\documentclass[preprint,showpacs,preprintnumbers,amsmath,amssymb]{revtex4}
\usepackage{graphicx}% Include figure files
\usepackage{dcolumn}% Align table columns on decimal point
\usepackage{bm}% bold math
\usepackage{color}
\usepackage{hhline}

\begin{document}
%\draft

\title{Enhancement of the spin transfer torque efficiency in magnetic STM junctions}

\author{Kriszti\'an Palot\'as}
\email{palotas@phy.bme.hu}
\affiliation{Budapest University of Technology and Economics,
Department of Theoretical Physics, Budafoki \'ut 8., H-1111 Budapest, Hungary}
\affiliation{Slovak Academy of Sciences, Institute of Physics, Department of Complex Physical Systems,
Center for Computational Materials Science, SK-84511 Bratislava, Slovakia}

\author{G\'abor M\'andi}
\affiliation{Budapest University of Technology and Economics,
Department of Theoretical Physics, Budafoki \'ut 8., H-1111 Budapest, Hungary}

\author{L\'aszl\'o Szunyogh}
\affiliation{Budapest University of Technology and Economics,
Department of Theoretical Physics and MTA-BME Condensed Matter Research Group,
Budafoki \'ut 8., H-1111 Budapest, Hungary}

\date{\today}

\begin{abstract}

We introduce a method for a combined calculation of charge and vector spin transport of elastically tunneling electrons in
magnetic scanning tunneling microscopy (STM). The method is based on the three-dimensional Wentzel-Kramers-Brillouin (3D-WKB)
approach combined with electronic structure calculations using first principles density functional theory.
As an application, we analyze the STM contrast inversion of the charge current above the Fe/W(110) surface depending on the
bias voltage, tip-sample distance and relative magnetization orientation between the sample and an iron tip.
For the spin transfer torque (STT) vector we find that its in-plane component is generally larger than the out-of-plane component,
and we identify a longitudinal spin current component, which, however, does not contribute to the torque.
Our results suggest that the torque-current relationship in magnetic STM junctions follows the power law rather than
a linear function. Consequently, we show that the ratio between the STT and the spin-polarized charge current is not constant,
and more importantly, it can be tuned by the bias voltage, tip-sample distance and magnetization rotation. We find that the
STT efficiency can be enhanced by about a factor of seven by selecting a proper bias voltage.
Thus, we demonstrate the possible enhancement of the STT efficiency in magnetic STM junctions, which can be exploited in
technological applications. We discuss our results in view of the indirect measurement of the STT above the Fe/W(110) surface
reported by Krause {\it et al.} in Phys.\ Rev.\ Lett.\ 107, 186601 (2011).

\end{abstract}


\maketitle

\section{Introduction}

Current induced magnetization switching in metallic spin valves and magnetic tunnel junctions (MTJs) is an intensive area of
research for its possible applications in future spintronic/magnetic devices \cite{ralph08}. During electronic transport,
inseparable from the charge current the electrons also transfer spins. If electron transfer occurs between two ferromagnets
having a noncollinear magnetic alignment, the transmitted spin transfer torque (STT) is directly responsible for the
magnetization switching. The first direct measurement of the in-plane and out-of-plane components of the STT vector in a MTJ was
performed by Sankey {\it et al.} \cite{sankey08}. On the other hand, there is a number of different theoretical approaches
reported to calculate the STT vector in MTJs. A brief overview is given below.

Free-electron models have been employed by several authors \cite{slonczewski96,wilczynski08,manchon08}. They used the
Wentzel-Kramers-Brillouin (WKB) approximation for the propagated wave functions. Slonczewski worked out the theory of STT within
Bardeen's transfer Hamiltonian method \cite{slonczewski05}. Tight-binding models in combination with Keldysh nonequilibrium
Green's function formalism have been extensively used \cite{theodonis06,kalitsov09,kalitsov13}. Very recently, the effects of
magnetic insulating \cite{pauyac14} and ferroelectric \cite{useinov15} barriers on the STT have been investigated by the
tight-binding method. A scattering theory of STT has been proposed by Xiao {\it et al.} \cite{xiao08}. A fully {\it{ab initio}}
method based on the Korringa-Kohn-Rostoker multiple scattering theory has been provided by Heiliger and Stiles \cite{heiliger08}.
As of particular technological importance, they studied the Fe/MgO/Fe MTJ, and found quantitative agreement with the
experimental STT results of Sankey {\it et al.} \cite{sankey08}. Jia {\it et al.} studied the same Fe/MgO/Fe MTJ by a
first-principles-based transport method and considered the effect of disorder scattering in MgO \cite{jia11}.

The listed theoretical models consider symmetric or asymmetric planar junction geometries to address STT effects in MTJs.
However, there is a growing interest to obtain information on local STT properties in high spatial resolution in asymmetric MTJs.
Spin-polarized scanning tunneling microscopy (SP-STM) proved to be extremely useful to provide local information on
spin-polarized charge transport properties in a wide variety of magnetic surfaces \cite{wiesendanger09review}.
As an important recent application, local current (and thus STT) pulses have been used to create and annihilate individual
magnetic skyrmion structures in a thin film system \cite{romming13}.
Based on above, we propose the generalization of high resolution STM charge transport theories \cite{hofer03rmp} to include the
description of vector spin transport in STM junctions. In this work, we present a combined theory for charge and vector spin
transport in magnetic STM junctions within the three-dimensional (3D) WKB electron tunneling framework, including electronic
structures of both the surface and the tip calculated by first principles methods. We investigate the Fe/W(110) surface motivated
by a recent indirect local STT measurement performed by SP-STM \cite{krause11}. Based on our results, we propose that the
STT efficiency in this system can be enhanced by about a factor of seven by selecting a proper bias voltage.
Moreover, given the exponentially decaying electron transmission in the MTJ with increasing tip-sample distance, we find that the
torque-current relationship follows the power law rather than a linear function.

The paper is organized as follows: The 3D-WKB theoretical model of combined charge and vector spin transport in magnetic STM is
presented in section \ref{sec_spstm}. Computational details of the electronic structure calculations are given in section
\ref{sec_comp}. We investigate the electron charge and spin transport in a magnetic tunnel junction consisting of one monolayer
Fe on the W(110) surface in combination with a Fe(001) tip in section \ref{sec_res}. We focus on the STM contrast formation of
the spin-polarized charge current and its determining factors in section \ref{sec_stm}, and study vector spin transport
characteristics by analyzing the spin transfer torque vector and the longitudinal spin current and their relation
to the spin-polarized charge current in section \ref{sec_torque}. Our conclusions are found in section \ref{sec_conc}.

\section{3D-WKB model of combined charge and vector spin transport in STM}
\label{sec_spstm}

Based on atom-superposition techniques \cite{tersoff85,yang02,smith04,heinze06}, the 3D-WKB electron tunneling theory
\cite{palotas11sts,palotas11stm,palotas12sts,palotas12orb} implemented in the 3D-WKB-STM code \cite{palotas14fop} became an
established method for the simulation of (spin-polarized) scanning tunneling microscopy and spectroscopy.
The model assumes that electron tunneling occurs through a tip apex atom, and individual transitions between this
tip apex atom and a suitable number of sample surface atoms, each described by the one-dimensional (1D) WKB approximation,
are superimposed. Orbital-dependent tunneling is included in the model by a modified transmission function \cite{palotas12orb}.
Since the 3D geometry of the tunnel junction is considered, the method is, in effect, a 3D-WKB atom-superposition approach.
Its main advantages are computational efficiency, see, e.g., Refs.\ \cite{palotas12orb,mandi15tipstat}, and an enhanced parameter
space for modeling tip geometries, i.e., arbitrary relative tip-sample orientations \cite{mandi13tiprot,mandi14rothopg}.
The 3D-WKB method has been applied to spin-polarized STM, and an analysis of the interplay of orbital-dependent and
spin-polarization effects was reported on the STM image contrast formation above the Fe(110) surface \cite{mandi14fe}.

For the combined 3D-WKB model for charge and vector spin transport, we adopt the orbital- and spin-dependent formalism
introduced in Ref.\ \cite{mandi14fe}.
The electronic structure of the magnetic surface and the tip is included in the model by taking the atom-projected electron
density of states (PDOS) obtained by {\it{ab initio}} electronic structure calculations.
The orbital-decomposition of the PDOS is necessary for the description of the orbital-dependent electron tunneling
\cite{palotas12orb}, and both the charge PDOS and the magnetization PDOS vector are essential for spin-dependent tunneling
\cite{palotas11stm}.
The energy-dependent orbital-decomposed charge PDOS of the $a$th sample surface ($S$) atom with orbital symmetry $\sigma$ and the
tip ($T$) apex atom with orbital symmetry $\tau$ are denoted by $n_{S\sigma}^a(E)$ and $n_{T\tau}(E)$, respectively.
In our model $\sigma,\tau\in\{s,p_y,p_z,p_x,d_{xy},d_{yz},d_{3z^2-r^2},d_{xz},d_{x^2-y^2}\}$ atomic orbitals are considered
but this can be extended to treat $f$ electron orbitals as well \cite{nita14}. Similarly to the charge PDOS,
$\mathbf{m}_{S\sigma}^a(E)$ and $\mathbf{m}_{T\tau}(E)$ denote the corresponding energy-dependent orbital-decomposed magnetization
PDOS vectors.

In the simplest case there are fixed spin quantization axes of the sample and the tip that can be described by energy-independent
$\mathbf{e}_S$ and $\mathbf{e}_T$ unit vectors, respectively. We focus on this case in the present work, allowing an arbitrary
$\phi$ angle between $\mathbf{e}_S$ and $\mathbf{e}_T$. The corresponding PDOS can be calculated from the spin-up
($n^{\uparrow}$) and spin-down ($n^{\downarrow}$) contributions obtained from first principles collinear magnetic calculations
for the sample surface and the tip separately \cite{palotas11stm}: The orbital-decomposed charge PDOS are
\begin{eqnarray}
n_{S\sigma}^a(E)&=&n^{a,\uparrow}_{S\sigma}(E)+n^{a,\downarrow}_{S\sigma}(E),\nonumber\\
n_{T\tau}(E)&=&n^{\uparrow}_{T\tau}(E)+n^{\downarrow}_{T\tau}(E),
\end{eqnarray}
and the orbital-decomposed magnetization PDOS vectors are
\begin{eqnarray}
\mathbf{m}_{S\sigma}^a(E)&=&m_{S\sigma}^a(E)\mathbf{e}_S=\left[n^{a,\uparrow}_{S\sigma}(E)-n^{a,\downarrow}_{S\sigma}(E)\right]\mathbf{e}_S,\nonumber\\
\mathbf{m}_{T\tau}(E)&=&m_{T\tau}(E)\mathbf{e}_T=\left[n^{\uparrow}_{T\tau}(E)-n^{\downarrow}_{T\tau}(E)\right]\mathbf{e}_T.
\end{eqnarray}
The total charge PDOS of the $a$th sample surface atom and the tip apex atom is the sum of the orbital-decomposed contributions:
\begin{eqnarray}
n_S^a(E)&=&\sum_{\sigma}n_{S\sigma}^a(E),\nonumber\\
n_T(E)&=&\sum_{\tau}n_{T\tau}(E).
\end{eqnarray}
Similarly, the total magnetization PDOS vectors can be obtained as the sum of the orbital-decomposed contributions:
\begin{eqnarray}
\mathbf{m}_S^a(E)&=&\sum_{\sigma}\mathbf{m}_{S\sigma}^a(E)=\left[\sum_{\sigma}m_{S\sigma}^a(E)\right]\mathbf{e}_S,\nonumber\\
\mathbf{m}_T(E)&=&\sum_{\tau}\mathbf{m}_{T\tau}(E)=\left[\sum_{\tau}m_{T\tau}(E)\right]\mathbf{e}_T.
\end{eqnarray}
Note that a similar decomposition of the Green's functions was reported within the linear combination of atomic orbitals (LCAO)
framework in Ref.\ \cite{mingo96}.

Let us define the following energy-dependent orbital-decomposed density matrices in spin space for the $a$th sample surface atom
and the tip apex atom \cite{hofer03magn,palotas11stm}:
\begin{eqnarray}
\underline{\underline{\rho}}_{S\sigma}^a(E)&=&n_{S\sigma}^a(E)\underline{\underline{I}}+\mathbf{m}_{S\sigma}^a(E)\cdot\underline{\underline{\bm{\sigma}}},\nonumber\\
\underline{\underline{\rho}}_{T\tau}(E)&=&n_{T\tau}(E)\underline{\underline{I}}+\mathbf{m}_{T\tau}(E)\cdot\underline{\underline{\bm{\sigma}}}.
\label{Eq_rho}
\end{eqnarray}
Here, $\underline{\underline{I}}$ is the $2\times 2$ unit matrix and $\underline{\underline{\bm{\sigma}}}=
\left(\underline{\underline{\sigma}}_x,\underline{\underline{\sigma}}_y,\underline{\underline{\sigma}}_z\right)$ is the
Pauli matrix vector. The main advantage of using the density matrix formalism is that electronic and spin
structures calculated either relativistically [with spin-orbit coupling included, general $\mathbf{m}_{S,T}(E)$] or
nonrelativistically [$\mathbf{m}_{S,T}(E)=m_{S,T}(E)\mathbf{e}_{S,T}$, used in the present work] can be treated within the same
theoretical framework \cite{palotas11stm}. The charge ($\tilde{I}$) and vector spin ($\tilde{\mathbf{T}}$) transport of the
electron tunneling transition between the $a$th sample surface atom and the tip apex atom at energy $E$ can be represented by
the following traces in spin space:
\begin{eqnarray}
\tilde{I}^a_{\sigma\tau}(E)&=&\frac{1}{2}Tr\left(\underline{\underline{\rho}}_{S\sigma}^a(E)\underline{\underline{\rho}}_{T\tau}(E)\underline{\underline{I}}\right)\nonumber\\
% &=&n_{S\sigma}^a(E)n_{T\tau}(E)+\mathbf{m}_{S\sigma}^a(E)\cdot\mathbf{m}_{T\tau}(E)\nonumber\\
&=&n_{S\sigma}^a(E)n_{T\tau}(E)+m_{S\sigma}^a(E)m_{T\tau}(E)\left[\mathbf{e}_S\cdot\mathbf{e}_T\right],\nonumber\\
\tilde{\mathbf{T}}^a_{\sigma\tau}(E)&=&\frac{1}{2}Tr\left(\underline{\underline{\rho}}_{S\sigma}^a(E)\underline{\underline{\rho}}_{T\tau}(E)\underline{\underline{\bm{\sigma}}}\right)\nonumber\\
% &=&\left[n_{S\sigma}^a(E)\mathbf{m}_{T\tau}(E)+\mathbf{m}_{S\sigma}^a(E)n_{T\tau}(E)\right]+i\left[\mathbf{m}_{S\sigma}^a(E)\times\mathbf{m}_{T\tau}(E)\right]\nonumber\\
&=&\left[n_{S\sigma}^a(E)m_{T\tau}(E)\mathbf{e}_T+m_{S\sigma}^a(E)n_{T\tau}(E)\mathbf{e}_S\right]+im_{S\sigma}^a(E)m_{T\tau}(E)\left[\mathbf{e}_S\times\mathbf{e}_T\right].
\label{Eq_chsp}
\end{eqnarray}
Since $\mathbf{e}_S\cdot\mathbf{e}_T=\cos(\phi)$ with $\phi$ the angle between the spin quantization axes of the sample surface
and the tip, the formula of the charge transport (conductance) is formally consistent with the spin-polarized Tersoff-Hamann model
\cite{wortmann01}, except the fact that it explicitly includes the electronic structure of the tip apex
\cite{palotas11stm,mandi14fe}:
\begin{equation}
\tilde{I}^a_{\sigma\tau}(E)=n_{S\sigma}^a(E)n_{T\tau}(E)+m_{S\sigma}^a(E)m_{T\tau}(E)\cos(\phi),
\end{equation}
where the first and second term is respectively the non-spin-polarized and spin-polarized part of the charge current contribution.
On the other hand, the in-plane and out-of-plane components of the spin transport formula are:
\begin{eqnarray}
\tilde{\mathbf{T}}^{a,in-pl.}_{\sigma\tau}(E)&=&\mathrm{Re}\left\{\tilde{\mathbf{T}}^a_{\sigma\tau}(E)\right\}=n_{S\sigma}^a(E)m_{T\tau}(E)\mathbf{e}_T+m_{S\sigma}^a(E)n_{T\tau}(E)\mathbf{e}_S,\nonumber\\
\tilde{\mathbf{T}}^{a,out-pl.}_{\sigma\tau}(E)&=&\mathrm{Im}\left\{\tilde{\mathbf{T}}^a_{\sigma\tau}(E)\right\}=m_{S\sigma}^a(E)m_{T\tau}(E)\sin(\phi)\mathbf{e}_{\{\mathbf{e}_S\times\mathbf{e}_T\}}.
\end{eqnarray}
Here, the in-plane component $\tilde{\mathbf{T}}^{in-pl.}$ lies in the plane spanned by the $\mathbf{e}_S$ and $\mathbf{e}_T$
vectors. Its direction is, however, not necessarily perpendicular to the magnetic moment the torkance is acting on, it rather
depends on the electronic structures of both the surface ($n_S,m_S$) and the tip ($n_T,m_T$) sides of the STM junction.
This implies that $\tilde{\mathbf{T}}^{in-pl.}$ can be decomposed to a longitudinal spin conductance part
(parallel to the magnetic moment direction $\mathbf{e}_j$) and to a spin torkance part (perpendicular to the magnetic moment
direction $\mathbf{e}_j$) \cite{xiao08}:
$\tilde{\mathbf{T}}^{a,in-pl.}_{\sigma\tau}(E)=\tilde{\mathbf{T}}^{a,long-j}_{\sigma\tau}(E)+\tilde{\mathbf{T}}^{a,j\parallel}_{\sigma\tau}(E)$,
where $\parallel$ denotes that this in-plane (Slonczewski) torkance vector is in the $\mathbf{e}_S$-$\mathbf{e}_T$ plane.
Here, the two components are:
\begin{eqnarray}
\tilde{\mathbf{T}}^{a,long-j}_{\sigma\tau}(E)&=&\left[\tilde{\mathbf{T}}^{a,in-pl.}_{\sigma\tau}(E)\cdot\mathbf{e}_j\right]\mathbf{e}_j=T^{a,long-j}_{\sigma\tau}(E)\mathbf{e}_j,\nonumber\\
\tilde{\mathbf{T}}^{a,j\parallel}_{\sigma\tau}(E)&=&\tilde{\mathbf{T}}^{a,in-pl.}_{\sigma\tau}(E)-T^{a,long-j}_{\sigma\tau}(E)\mathbf{e}_j,
\end{eqnarray}
with $j\in\{S,T\}$ depending on the direction of spin transport, i.e., $j=S$ for tip $\rightarrow$ sample and $j=T$ for
sample $\rightarrow$ tip tunneling. The longitudinal spin conductance term used for $\tilde{\mathbf{T}}^{long-j}$ is
motivated by the limiting case of $\mathbf{e}_S=\mathbf{e}_T$ ($\phi=0$), where
$\tilde{T}^{long-S=T}\propto n_S^{\uparrow}n_T^{\uparrow}-n_S^{\downarrow}n_T^{\downarrow}$.

Since $\mathbf{e}_S\times\mathbf{e}_T=\sin(\phi)\mathbf{e}_{\{\mathbf{e}_S\times\mathbf{e}_T\}}$,
$\tilde{\mathbf{T}}^{out-pl.}$ is perpendicular to the $\mathbf{e}_S$-$\mathbf{e}_T$ plane and in line with the
$\mathbf{e}_{\{\mathbf{e}_S\times\mathbf{e}_T\}}$ normal-to-plane unit vector. Its size depends on the magnetization PDOS
$m_S,m_T$ and on the $\phi$ angle. Since this vector is definitely perpendicular to both magnetic moments,
we identify it as the out-of-plane (field-like) torkance vector \cite{xiao08}:
$\tilde{\mathbf{T}}^{a,out-pl.}_{\sigma\tau}(E)=\tilde{\mathbf{T}}^{a,\perp}_{\sigma\tau}(E)$.
From the above formalism it is clear that this out-of-plane torkance corresponds to conductance-induced torkance
solely as a result of electron tunneling through the vacuum barrier and does not include contributions related to the
interlayer exchange coupling in planar junctions \cite{slonczewski96,kalitsov13} that result in the so-called equilibrium torque
\cite{theodonis06}, which is non-zero even at zero bias voltage ($T_{\perp}(0)\ne 0$). Therefore, throughout this work,
the out-of-plane torque corresponds to purely current-induced torque, $T^{\perp}(V)=T_{\perp}(V)-T_{\perp}(0)$ using the notation
of Ref.\ \cite{theodonis06}.

The total tunneling charge current, the components of the spin transfer torque (STT) vector and the longitudinal
spin current are calculated from the charge conductance, the components of the spin torkance vector and the
longitudinal spin conductance by the superposition of atomic contributions from the sample surface
(sum over $a$) and the superposition of transitions from all atomic orbital combinations between the sample and the tip
(sum over $\sigma$ and $\tau$) integrated in an energy window corresponding to the bias voltage $V$. Assuming elastic electron
tunneling at temperature $T=0$ K, the charge current, the STT vector components, the longitudinal spin current,
the STT vector and the magnitude of the STT at the tip apex position $\mathbf{R}_{TIP}$ and bias voltage $V$ are given by
\begin{eqnarray}
I\left(\mathbf{R}_{TIP},V\right)&=&\sum_a\sum_{\sigma,\tau}I^a_{\sigma\tau}\left(\mathbf{R}_{TIP},V\right),\nonumber\\
\mathbf{T}^{\perp}\left(\mathbf{R}_{TIP},V\right)&=&\sum_a\sum_{\sigma,\tau}\mathbf{T}^{a,\perp}_{\sigma\tau}\left(\mathbf{R}_{TIP},V\right),\nonumber\\
\mathbf{T}^{j\parallel}\left(\mathbf{R}_{TIP},V\right)&=&\sum_a\sum_{\sigma,\tau}\mathbf{T}^{a,j\parallel}_{\sigma\tau}\left(\mathbf{R}_{TIP},V\right),\nonumber\\
T^{long-j}\left(\mathbf{R}_{TIP},V\right)&=&\sum_a\sum_{\sigma,\tau}T^{a,long-j}_{\sigma\tau}\left(\mathbf{R}_{TIP},V\right),\nonumber\\
\mathbf{T}^j\left(\mathbf{R}_{TIP},V\right)&=&\mathbf{T}^{j\parallel}\left(\mathbf{R}_{TIP},V\right)+\mathbf{T}^{\perp}\left(\mathbf{R}_{TIP},V\right),\nonumber\\
T^j\left(\mathbf{R}_{TIP},V\right)&=&\left[\left|\mathbf{T}^{j\parallel}\left(\mathbf{R}_{TIP},V\right)\right|^2+\left|\mathbf{T}^{\perp}\left(\mathbf{R}_{TIP},V\right)\right|^2\right]^{1/2},
\label{Eq_current}
\end{eqnarray}
with $j\in\{S,T\}$ depending on the direction of spin transport, i.e., $j=S$ for tip $\rightarrow$ sample and
$j=T$ for sample $\rightarrow$ tip tunneling. Therefore, $j$ naturally labels the side (magnetic surface or tip), where the torque
is acting on.
One particular contribution can be calculated as an integral in an energy window corresponding to the bias voltage $V$ as
\begin{eqnarray}
I^a_{\sigma\tau}\left(\mathbf{R}_{TIP},V\right)&=&\epsilon^2\frac{e^2}{h}\int_0^V\hat{T}_{\sigma\tau}\left(E_F^S+eU,V,\mathbf{d}_a\right)\nonumber\\
&\times&\left[n_{S\sigma}^a\left(E_F^S+eU\right)n_{T\tau}\left(E_F^T+eU-eV\right)\right.\nonumber\\
&+&\left. m_{S\sigma}^a\left(E_F^S+eU\right)m_{T\tau}\left(E_F^T+eU-eV\right)\cos(\phi)\right]dU,\nonumber\\
\mathbf{T}^{a,\perp}_{\sigma\tau}\left(\mathbf{R}_{TIP},V\right)&=&\epsilon^2 e\int_0^V\hat{T}_{\sigma\tau}\left(E_F^S+eU,V,\mathbf{d}_a\right)\nonumber\\
&\times&m_{S\sigma}^a\left(E_F^S+eU\right)m_{T\tau}\left(E_F^T+eU-eV\right)\sin(\phi)\mathbf{e}_{\{\mathbf{e}_S\times\mathbf{e}_T\}}dU,\nonumber\\
\mathbf{T}^{a,S\parallel}_{\sigma\tau}\left(\mathbf{R}_{TIP},V\right)&=&\epsilon^2 e\int_0^V\hat{T}_{\sigma\tau}\left(E_F^S+eU,V,\mathbf{d}_a\right)\nonumber\\
&\times&n_{S\sigma}^a\left(E_F^S+eU\right)m_{T\tau}\left(E_F^T+eU-eV\right)\left[\mathbf{e}_T-\mathbf{e}_S\cos(\phi)\right]dU,\nonumber\\
\mathbf{T}^{a,T\parallel}_{\sigma\tau}\left(\mathbf{R}_{TIP},V\right)&=&\epsilon^2 e\int_0^V\hat{T}_{\sigma\tau}\left(E_F^S+eU,V,\mathbf{d}_a\right)\nonumber\\
&\times&m_{S\sigma}^a\left(E_F^S+eU\right)n_{T\tau}\left(E_F^T+eU-eV\right)\left[\mathbf{e}_S-\mathbf{e}_T\cos(\phi)\right]dU,\nonumber\\
T^{a,long-S}_{\sigma\tau}\left(\mathbf{R}_{TIP},V\right)&=&\epsilon^2 e\int_0^V\hat{T}_{\sigma\tau}\left(E_F^S+eU,V,\mathbf{d}_a\right)\nonumber\\
&\times&\left[n_{S\sigma}^a\left(E_F^S+eU\right)m_{T\tau}\left(E_F^T+eU-eV\right)\cos(\phi)\right.\nonumber\\
&+&\left. m_{S\sigma}^a\left(E_F^S+eU\right)n_{T\tau}\left(E_F^T+eU-eV\right)\right]dU,\nonumber\\
T^{a,long-T}_{\sigma\tau}\left(\mathbf{R}_{TIP},V\right)&=&\epsilon^2 e\int_0^V\hat{T}_{\sigma\tau}\left(E_F^S+eU,V,\mathbf{d}_a\right)\nonumber\\
&\times&\left[n_{S\sigma}^a\left(E_F^S+eU\right)m_{T\tau}\left(E_F^T+eU-eV\right)\right.\nonumber\\
&+&\left. m_{S\sigma}^a\left(E_F^S+eU\right)n_{T\tau}\left(E_F^T+eU-eV\right)\cos(\phi)\right]dU.
\label{Eq_current_decomp}
\end{eqnarray}
Here, $e$ is the elementary charge, $h$ is the Planck constant, and $E_F^S$ and $E_F^T$ are the Fermi energies of the sample
surface and the tip, respectively. The $\epsilon^{2}e^{2}/h$ and $\epsilon^{2}e$ factors ensure the correct
dimensions of the quantities.
The value of $\epsilon$ has to be determined by comparing the simulation results of the charge current with experiments, or with
calculations using standard methods, e.g., the Bardeen approach \cite{bardeen61}. In our simulations $\epsilon=1$ eV was chosen
that gives comparable current values with those obtained by the Bardeen method \cite{palotas12orb} implemented in the BSKAN code
\cite{hofer03pssci,palotas05}. Note that the choice of $\epsilon$ has no qualitative influence on the reported results.
The formal equivalence of the in-plane torque vectors in Eq.\ (\ref{Eq_current_decomp}) with Eq.\ (6) of
Ref.\ \cite{theodonis06} is proven in Appendix.

In Eq.\ (\ref{Eq_current_decomp}), $\hat{T}_{\sigma\tau}\left(E,V,\mathbf{d}_a\right)$ is the orbital-dependent tunneling
transmission function that gives the probability of the electron charge/spin tunneling from the $\tau$ orbital of the tip apex
atom to the $\sigma$ orbital of the $a$th surface atom, or vice versa, depending on the sign of the bias voltage. The conventions
of tip $\rightarrow$ sample tunneling at positive bias voltage ($V>0$) and sample $\rightarrow$ tip tunneling at negative bias
($V<0$) are used. The STT is acting on the magnetic moments where the charge current flows, i.e., on the sample moments at
positive and on the tip moments at negative bias voltage.
The transmission probability depends on the energy of the electron ($E$), the bias voltage ($V$), and the relative position of the
tip apex atom and the $a$th sample surface atom ($\mathbf{d}_a=\mathbf{R}_{TIP}-\mathbf{R}_a$).
The following form for the transmission function is considered \cite{mandi13tiprot}:
\begin{equation}
\label{Eq_Transmission}
\hat{T}_{\sigma\tau}\left(E_F^S+eU,V,\mathbf{d}_a\right)=\exp\{-2\kappa(U,V)|\mathbf{d}_a|\}\chi_{\sigma}^2(\theta_a,\varphi_a)\chi_{\tau}^2(\theta_a',\varphi_a').
\end{equation}
Here, the exponential factor corresponds to an orbital-independent transmission, where all electron states are considered as
exponentially decaying spherical states \cite{tersoff85,heinze06,tersoff83}. This factor depends on the distance between the
tip apex atom and the $a$th surface atom, $|\mathbf{d}_a|$, and on the 1D-WKB vacuum decay of electron states,
\begin{equation}
\label{Eq_kappa}
\kappa(U,V)=\frac{1}{\hbar}\sqrt{2m\left(\frac{\Phi_S+\Phi_T+eV}{2}-eU\right)}.
\end{equation}
For using this $\kappa$ an effective rectangular potential barrier in the vacuum between the sample and the tip is assumed.
$\Phi_S$ and $\Phi_T$ are the electron work functions of the sample surface and the tip, respectively, $m$ is the electron's
mass and $\hbar$ the reduced Planck constant. The remaining factors of Eq.\ (\ref{Eq_Transmission}) are responsible for the
orbital dependence of the transmission. They modify the exponentially decaying part according to the real-space shape of the
electron orbitals involved in the tunneling, i.e., the angular dependence of the electron densities of the atomic orbitals of
the surface and the tip is taken into account as the square of the real spherical harmonics
$\chi_{\sigma}(\theta_a,\varphi_a)$ and $\chi_{\tau}(\theta_a',\varphi_a')$, respectively. The distinction between the local
coordinate systems of the sample surface (without primes) and the tip apex atom (denoted by primes) enables the consideration of
arbitrary relative tip-sample geometrical orientations within the orbital-dependent 3D-WKB approach
\cite{mandi13tiprot,mandi14rothopg,mandi15tipstat}.
The polar and azimuthal angles $(\theta_a^{(')},\varphi_a^{(')})$ given in both real spherical harmonics in
Eq.\ (\ref{Eq_Transmission}) correspond to the tunneling direction, i.e., the line connecting the $a$th surface atom and the
tip apex atom, as viewed from their local coordinate systems, and they have to be determined from the actual tip-sample geometry.

It is important to highlight that the magnitude of both STT vector components, and thus the magnitude of the STT,
are proportional to $|\sin(\phi)|$. On the other hand, the spin-polarized part of the charge current is proportional to
$\cos(\phi)$. Therefore, if $\mathbf{e}_S$ and $\mathbf{e}_T$ are in line ($\phi=0$ or $\phi=\pm\pi$) then both torque vector
components and the STT vanish and the spin-polarized part of the charge current is largest in magnitude.

\section{Computational details}
\label{sec_comp}

We have performed geometry optimization and electronic structure calculations within the generalized gradient approximation (GGA)
of the density functional theory (DFT) implemented in the Vienna Ab-initio Simulation Package (VASP) \cite{VASP2,VASP3,hafner08}.
A plane wave basis set for electronic wave function expansion in combination with the projector augmented wave (PAW) method
\cite{kresse99} has been applied. We have adopted the parametrization of the exchange-correlation functional by Perdew and Wang
(PW91) \cite{pw91}. The electronic structures of the Fe/W(110) surface and the Fe(001) tip have been calculated separately.
It should be noted that spin-orbit coupling effects have been neglected in our DFT calculations, and
$\textbf{m}_{S,T}(E)=m_{S,T}(E)\textbf{e}_{S,T}$ in Eq.\ (\ref{Eq_rho}).

We model the Fe/W(110) surface by a slab of nine atomic layers (8 tungsten and 1 iron layers) with the bulk lattice constant of
$a_{W}=3.165$ \AA. A separating vacuum region of 18 \AA\;width in the surface normal ($z$) direction between neighboring supercell
slabs is set up to minimize slab-slab interaction. After geometry optimization the Fe-W interlayer distance between the two
topmost layers is reduced by 14\%, and the underneath W-W interlayer distance is increased by 1\% in comparison to bulk tungsten.
The magnitude of the in-plane spin moment of the surface Fe atoms is 2.31 $\mu_B$ and the induced moments in the neighboring
W layer are -0.14 $\mu_B$ per tungsten atom (opposite in orientation compared to the surface Fe spin moments).
The structural parameters and magnetic properties of Fe/W(110) are in good agreement with Refs.\ \cite{qian99,rozsa15}.
The spin quantization axis ($\mathbf{e}_S$, magnetic easy axis) is pointing to the $[1\bar{1}0]$ crystallographic direction.
The average electron work function above the Fe/W(110) surface is $\Phi_S=4.73$ eV calculated from the local electrostatic
potential. We have used an $11\times 11\times 1$ Monkhorst-Pack \cite{monkhorst} k-point grid for obtaining the orbital-decomposed
projected charge and magnetization electron DOS onto the surface Fe atom, $n_{S\sigma}(E)$ and $m_{S\sigma}(E)$, respectively.

We model the STM tip as a blunt iron tip, where a single Fe apex atom is placed on the hollow position of the Fe(001) surface,
similarly as in Ref.\ \cite{ferriani10tip}. More details of the Fe(001) tip model can be found in Ref.\ \cite{mandi14fe}.
The orbital-decomposed electronic structure data projected to the tip apex atom, $n_{T\tau}(E)$ and $m_{T\tau}(E)$, have been
calculated using a $13\times 13\times 3$ Monkhorst-Pack k-point grid. The orientation of the tip spin quantization axis
$\mathbf{e}_T$ relative to $\mathbf{e}_S$ is arbitrarily changed and their angle is denoted by $\phi$.

For calculating the spin-polarized charge current and the spin transfer torque vector, we choose 112 iron atoms on the Fe/W(110)
surface for the atomic superposition \cite{mandi14fe}.

\section{Results and discussion}
\label{sec_res}

\begin{figure*}
\begin{tabular}{cc}
\includegraphics[width=0.5\textwidth,angle=0]{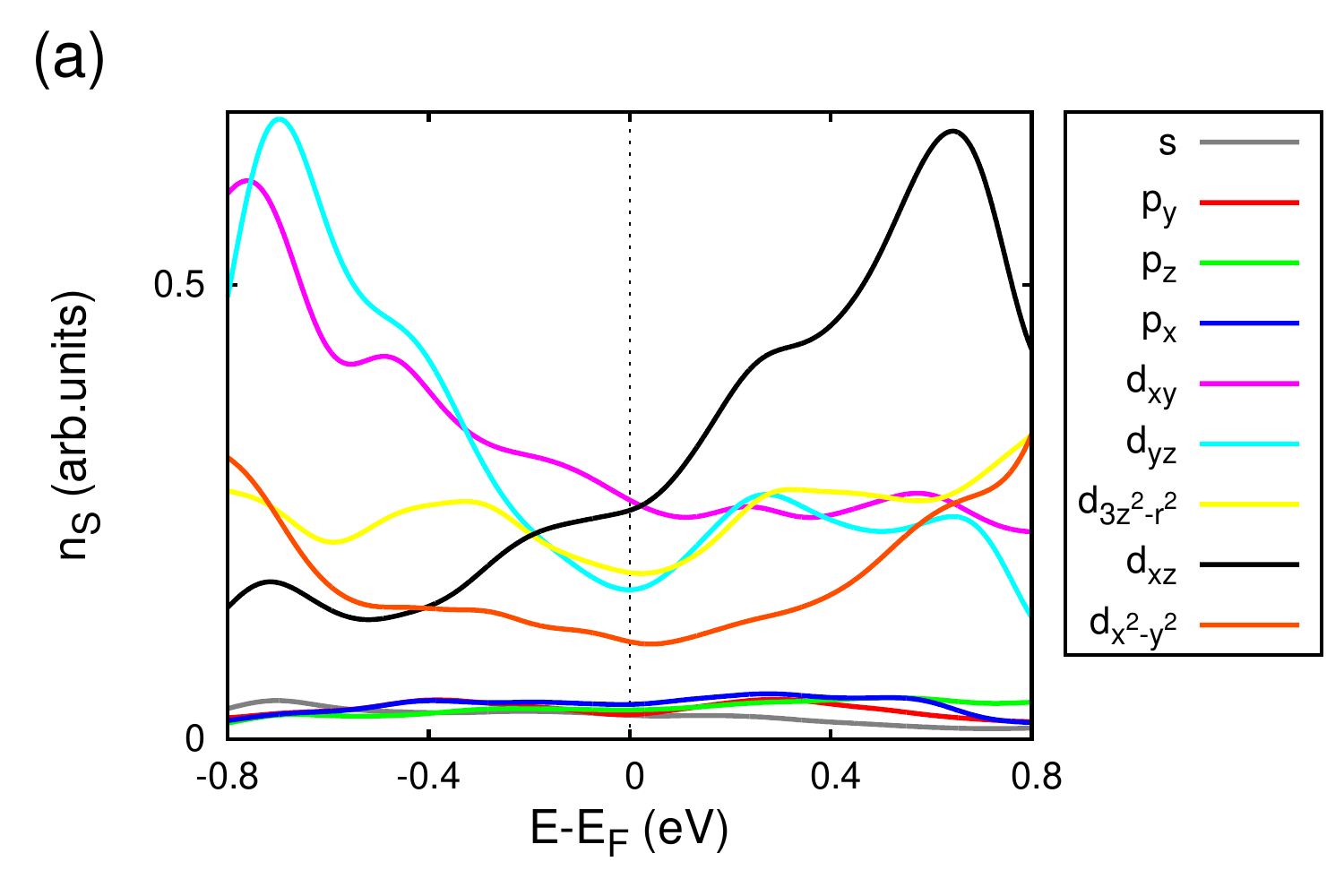}&
\includegraphics[width=0.5\textwidth,angle=0]{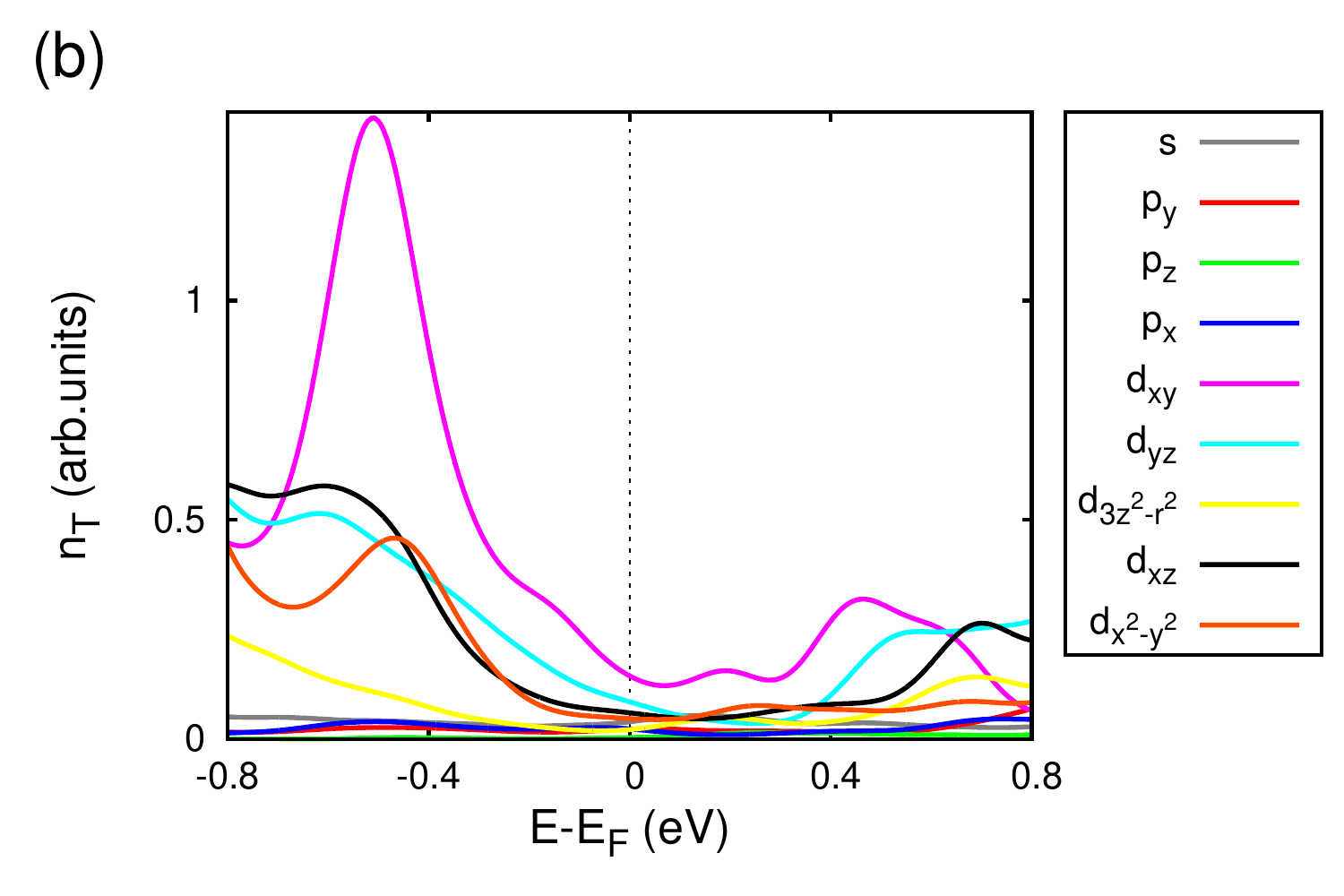}\tabularnewline
\includegraphics[width=0.5\textwidth,angle=0]{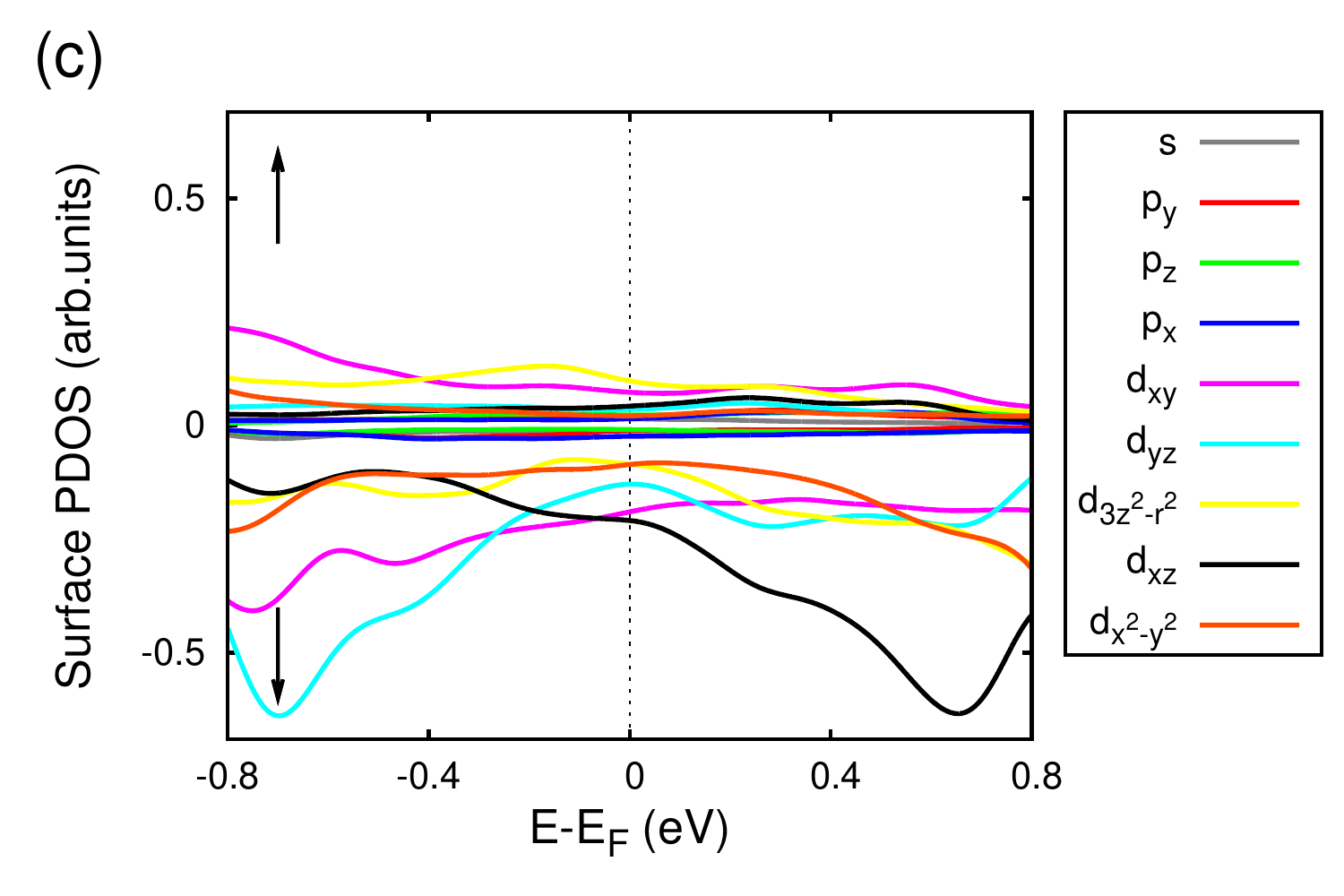}&
\includegraphics[width=0.5\textwidth,angle=0]{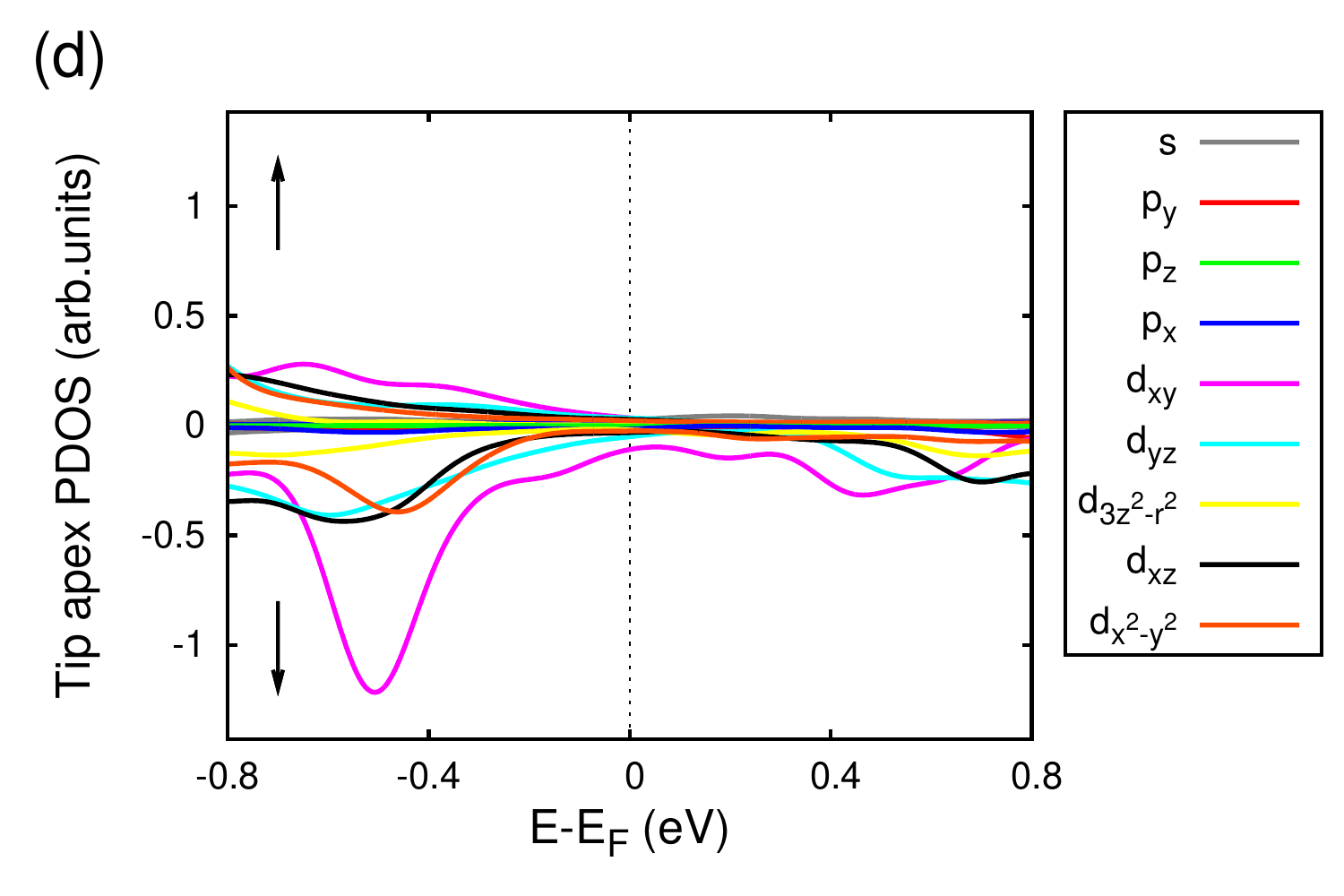}
\end{tabular}
\caption{\label{Fig1} (Color online) Orbital-decomposed projected electron density of states (PDOS) of the Fe surface atom in
Fe/W(110) and the iron tip apex atom.
(a) Fe/W(110) surface Fe charge PDOS: $n_{S\sigma}(E)$;
(b) tip apex Fe charge PDOS: $n_{T\tau}(E)$;
(c) Fe/W(110) surface Fe spin-resolved PDOS: $n_{S\sigma}^{\uparrow,\downarrow}(E)$;
(d) tip apex Fe spin-resolved PDOS: $n_{T\tau}^{\uparrow,\downarrow}(E)$.
Orbitals $\sigma,\tau\in\{s,p_y,p_z,p_x,d_{xy},d_{yz},d_{3z^2-r^2},d_{xz},d_{x^2-y^2}\}$ are indicated.
}
\end{figure*}

Figure \ref{Fig1} shows the energy-dependent orbital-decomposed charge PDOS ($n=n^{\uparrow}+n^{\downarrow}$) and spin-resolved
PDOS ($n^{\uparrow,\downarrow}$) functions of the Fe surface atom in Fe/W(110) and the Fe(001) tip apex atom. We find that the
$d$ partial PDOS is dominating over $s$ and $p$ partial PDOS, and the minority spin character is dominating over the majority spin
in the vicinity of the Fermi levels. The obtained results are in good agreement with the total (orbital-summed) Fe PDOS functions
reported in Refs.\ \cite{qian99,andersen06,ferriani10tip}, where the full potential linearized augmented plane wave (FLAPW) method
was employed in different computational implementations (Wien2k and FLEUR codes).

\subsection{Charge current STM contrast inversion}
\label{sec_stm}

As demonstrated for W(110) and Fe(110) surfaces in Refs.\ \cite{palotas12orb,mandi14fe}, the charge current difference between
tip positions above the top and hollow surface sites of BCC(110) surfaces is indicative for the corrugation character of a
constant-current STM image. The charge current difference at tip-sample distance $z$ and bias voltage $V$ is defined as
\begin{equation}
\label{Eq_deltaI}
\Delta I(z,V)=I^{top}(z,V)-I^{hollow}(z,V).
\end{equation}
Positive $\Delta I$ corresponds to an STM image with normal corrugation, where the atomic sites appear as protrusions.
Conversely, negative $\Delta I$ is related to an STM image with inverted contrast, i.e., anticorrugation, and atomic sites
appearing as depressions. The $\Delta I(z,V)=0$ contour gives the $(z,V)$ combinations where the corrugation inversion occurs.

\begin{figure*}
\includegraphics[width=0.8\textwidth,angle=0]{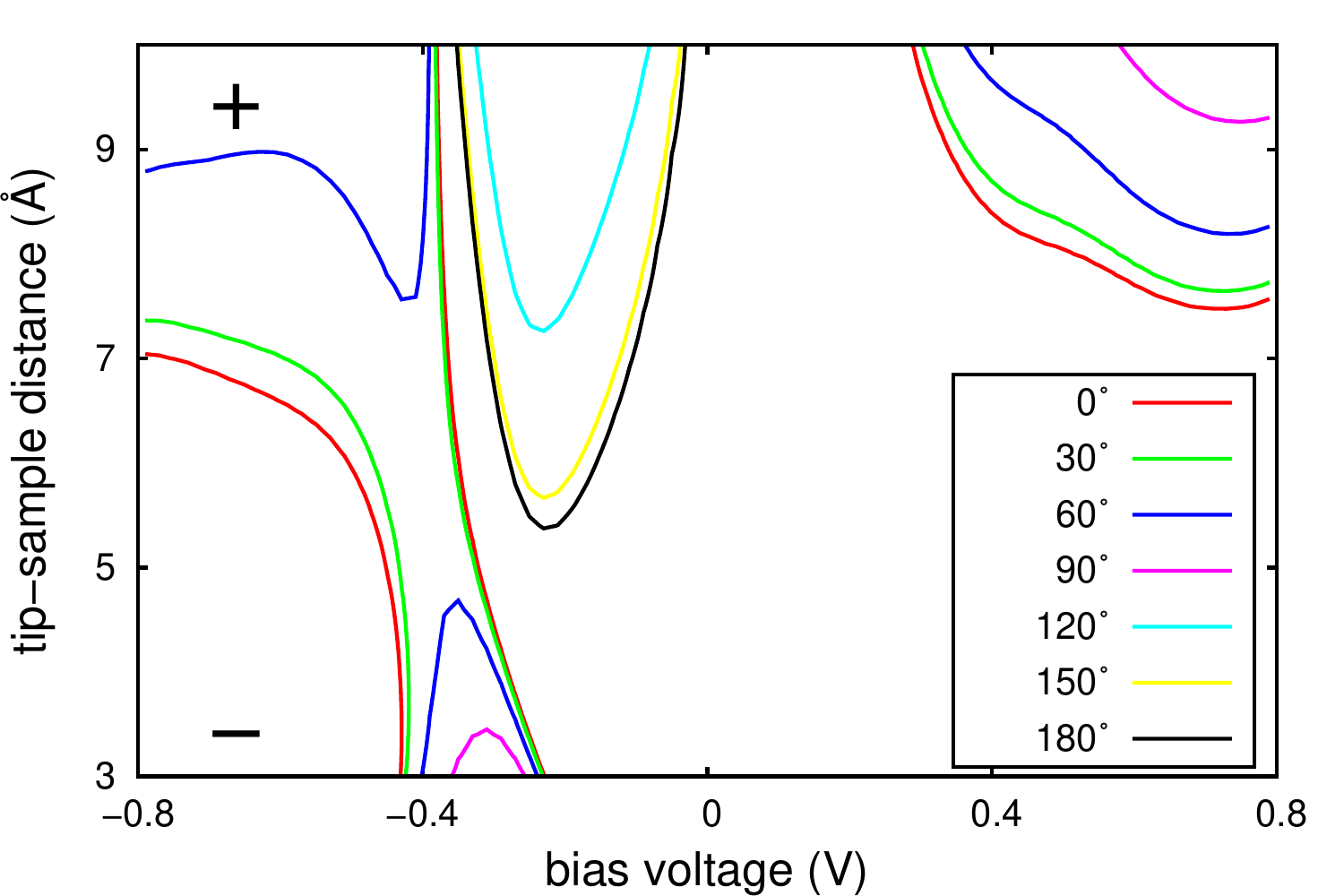}
\caption{\label{Fig2} (Color online) $\Delta I(z,V)=I^{top}(z,V)-I^{hollow}(z,V)=0$ contours indicative for the corrugation
inversion of STM images [see Eq.\ (\ref{Eq_deltaI}), and its meaning in the text] above the Fe/W(110) surface in
seven relative tip-sample magnetization orientations ($\phi$ between 0 and 180 degrees in $30^{\circ}$ step).
The sign of $\Delta I$ ($+$ or $-$) is shown in selected $(z,V)$ sections, and crossing the $\Delta I=0$ curve means inversion of
the sign, hence inversion of the spin-polarized charge current contrast in STM. Note that positive $\Delta I$ corresponds to
normal, and negative to inverted atomic contrast in STM images, for a demonstration see Refs.\ \cite{palotas12orb,mandi14fe}.
}
\end{figure*}

Figure \ref{Fig2} shows such zero current difference contours above the Fe/W(110) surface in the [-0.8 V, +0.8 V] bias voltage
and [3 \AA, 10 \AA] tip-sample distance ranges assuming seven relative tip-sample magnetization orientations.
We find that all considered $\phi$ angles result in the appearance of contrast inversions. Close to the surface
anticorrugation ($\Delta I<0$) is observed with the exception of the bias voltage at around -0.4 V for $\phi$ angles upto
90 degrees. Farther from the surface two important effects can be seen: (i) the $(z,V)$ regions with normal corrugation
($\Delta I>0$) above +0.3 V and below -0.4 V are shrinking with increasing the $\phi$ angle in the [$0^{\circ}$, $90^{\circ}$]
range, and (ii) a $(z,V)$ region with normal corrugation opens up centered at -0.2 V upon increasing the $\phi$ angle in the
[$120^{\circ}$, $180^{\circ}$] range.

For the understanding of these findings we recall that the complex interplay of orbital-dependent and spin-polarization
effects determine the STM corrugation character above a magnetic surface. While the orbitals with $m=0$ magnetic quantum number
($s$, $p_z$, $d_{3z^2-r^2}$) prefer normal corrugation, $m\ne 0$ states generally prefer anticorrugation
\cite{chen92,palotas12orb}. This electron orbital picture is modified by taking the weight of the contributing spin channels to
the spin-polarized electron tunneling in a magnetic STM into account \cite{mandi14fe}. Thus, based on the 3D-WKB model,
the STM contrast depending on the bias voltage and on the tip-sample distance can be explained as an interplay of the real-space
orbital shapes involved in the electron tunneling and their corresponding spin-resolved energy-dependent partial PDOS.

The simplest explanation of the STM contrast character can be given for the $\phi=90^{\circ}$ case, i.e., for perpendicular spin
quantization axes of the surface and the tip. This particular setup results in zero spin-polarized contribution to the tunneling
current because $\cos(\phi)=0$, and only orbital shape effects play a role in the tunneling without spin-polarization effects.
This is governed by the orbital-dependent transmission function in Eq.\ (\ref{Eq_Transmission}).
Fig.\ \ref{Fig1}(c) shows that the $d$ partial PDOS of the Fe/W(110) surface outweighs the $s$ and $p$ partial PDOS in the whole
energy range. Taking $d$ orbitals of the surface and following geometrical considerations, the leading current contribution is
expected from the $d_{3z^2-r^2}$ orbital of the underlying Fe atom and from the $d_{xy}$ orbital from the nearest neighbor Fe
atoms when the tip is placed above the surface top position ($I^{top}$). On the other hand, the main current contribution above
the surface hollow position ($I^{hollow}$) is expected from the four nearest neighbor Fe atoms and their $d_{xz}$ and $d_{yz}$
orbitals close to the surface and $d_{3z^2-r^2}$ orbital at larger tip-sample distances. Since Fig.\ \ref{Fig1}(a) shows that the
$d_{xz}$ partial PDOS is the largest above the surface Fermi level and $d_{yz}$ and $d_{xy}$ partial PDOS are the largest below
$E_F^S$, $I^{hollow}$ is expected to be larger than $I^{top}$, and the overall current difference $\Delta I=I^{top}-I^{hollow}$
will be negative in almost the entire studied $(z,V)$ range. Two exceptions are (i) close to -0.3 V and close to the surface,
where the overall effect of $d_{xy}$ and $d_{3z^2-r^2}$ partial PDOS outweighs that of $d_{yz}$, and (ii) above +0.6 V at large
tip-sample distance, where the geometry effect of the $d_{3z^2-r^2}$ orbital dominates, and in both $(z,V)$ regimes
$I^{top}>I^{hollow}$ and normal corrugation is found.

Different $\phi$ angles than $90^{\circ}$ result in the occurrence of spin-polarized charge current, and the above described
physical picture is modified by the contributing spin channels. The [$0^{\circ}$, $90^{\circ}$) $\phi$ angle range corresponds to
the dominance of the so-called ferromagnetic current
[$I_F=I(\phi=0^{\circ})\propto n_S^{\uparrow}n_T^{\uparrow}+n_S^{\downarrow}n_T^{\downarrow}$],
and in the $\phi=(90^{\circ}$, $180^{\circ}]$ interval the weight of the antiferromagnetic current
[$I_A=I(\phi=180^{\circ})\propto n_S^{\uparrow}n_T^{\downarrow}+n_S^{\downarrow}n_T^{\uparrow}$] is larger than that of $I_F$
\cite{hofer03magn}. Note that with the introduced notations $I(\phi=90^{\circ})=(I_F+I_A)/2$, i.e., an equal 0.5-0.5
weighting of $I_F$ and $I_A$ is achieved.

We find that both the shrinking of the $(z,V)$ regions showing normal corrugation above +0.3 V and below -0.4 V with increasing
the $\phi$ angle in the [$0^{\circ}$, $90^{\circ}$] range and the opening of the $(z,V)$ region with normal corrugation centered
at -0.2 V upon increasing the $\phi$ angle in the [$120^{\circ}$, $180^{\circ}$] range result from the increasing importance
of $I_A$, i.e., the mixture of the sample and tip majority and minority spin contributions. For instance, in the former case
the minority spin $d_{xy}$ PDOS peak of the tip located at 0.5 eV below its Fermi level [see Fig.\ \ref{Fig1}(d)] plays an
important role. The $d_{xy}$ tip orbital prefers normal corrugation due to its $xz$ and $yz$ nodal planes since above the hollow
position the anticorrugating $d_{xz,yz}$ surface orbitals do not contribute to the current resulting in lower $I^{hollow}$ than
$I^{top}$ \cite{palotas12orb}. On the other hand, in the latter case $d_{3z^2-r^2}$ and $d_{xy}$ orbital characters of PDOS
dominate between $[E_F^S-0.4$ eV, $E_F^S]$ in the majority spin channel of the Fe/W(110) surface. This combined with the minority
spin PDOS of the iron tip results in the opening of a normally corrugated $(z,V)$ region around -0.2 V.

The above findings clearly demonstrate the tunability of the atomic contrast inversion in STM images depending on the
contributing spin channels and consequently on the relative tip-sample magnetization orientation.

\subsection{Vector spin transport characteristics}
\label{sec_torque}

In this section we present a detailed analysis of the calculated vector spin transport components and their
relation to the spin-polarized charge current by varying different tunnel junction parameters, such as the relative tip-sample
magnetization orientation, the lateral and vertical positions of the STM tip and the bias voltage.

\begin{figure*}
\begin{tabular}{cc}
\includegraphics[width=0.5\textwidth,angle=0]{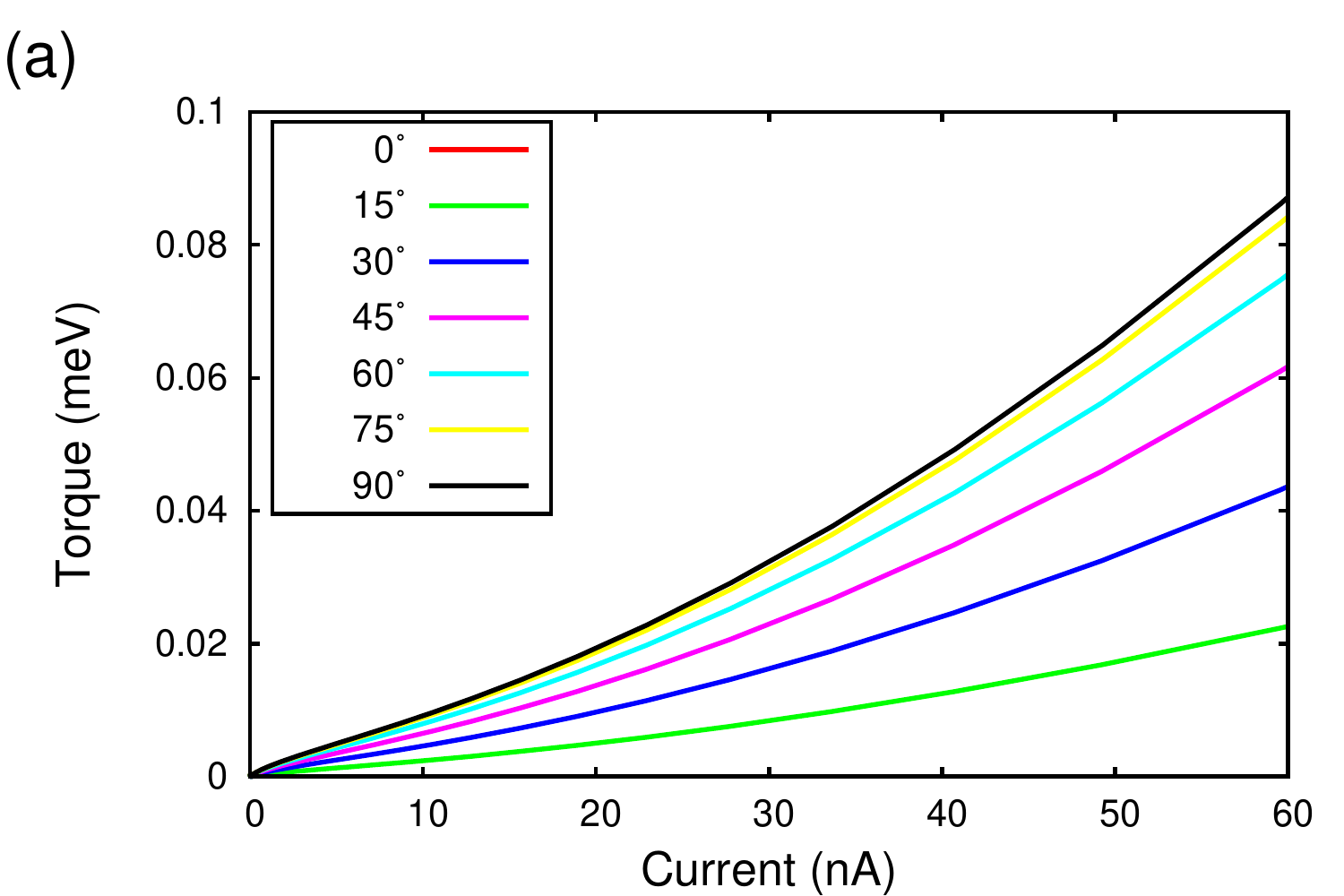}&
\includegraphics[width=0.5\textwidth,angle=0]{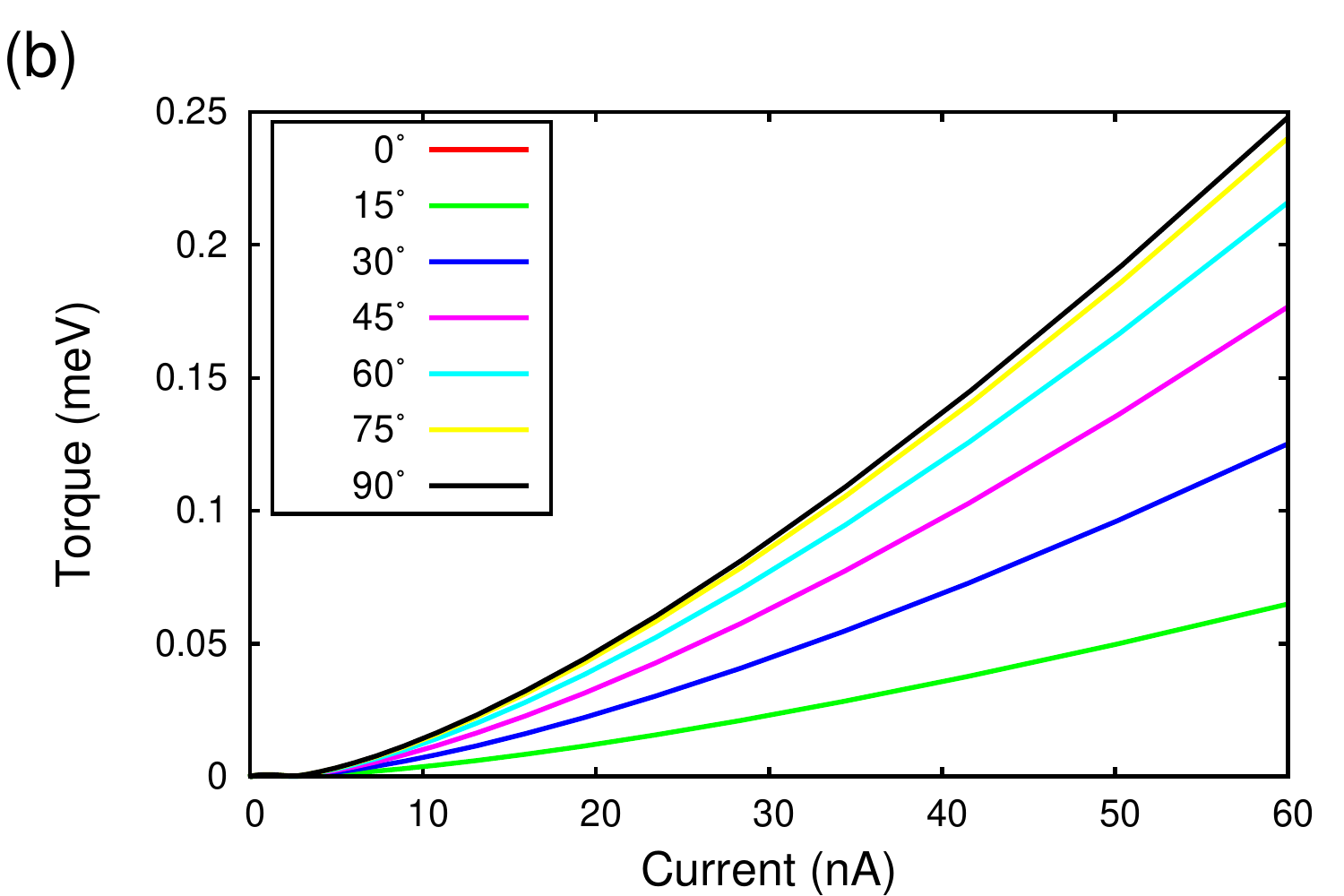}\tabularnewline
\includegraphics[width=0.5\textwidth,angle=0]{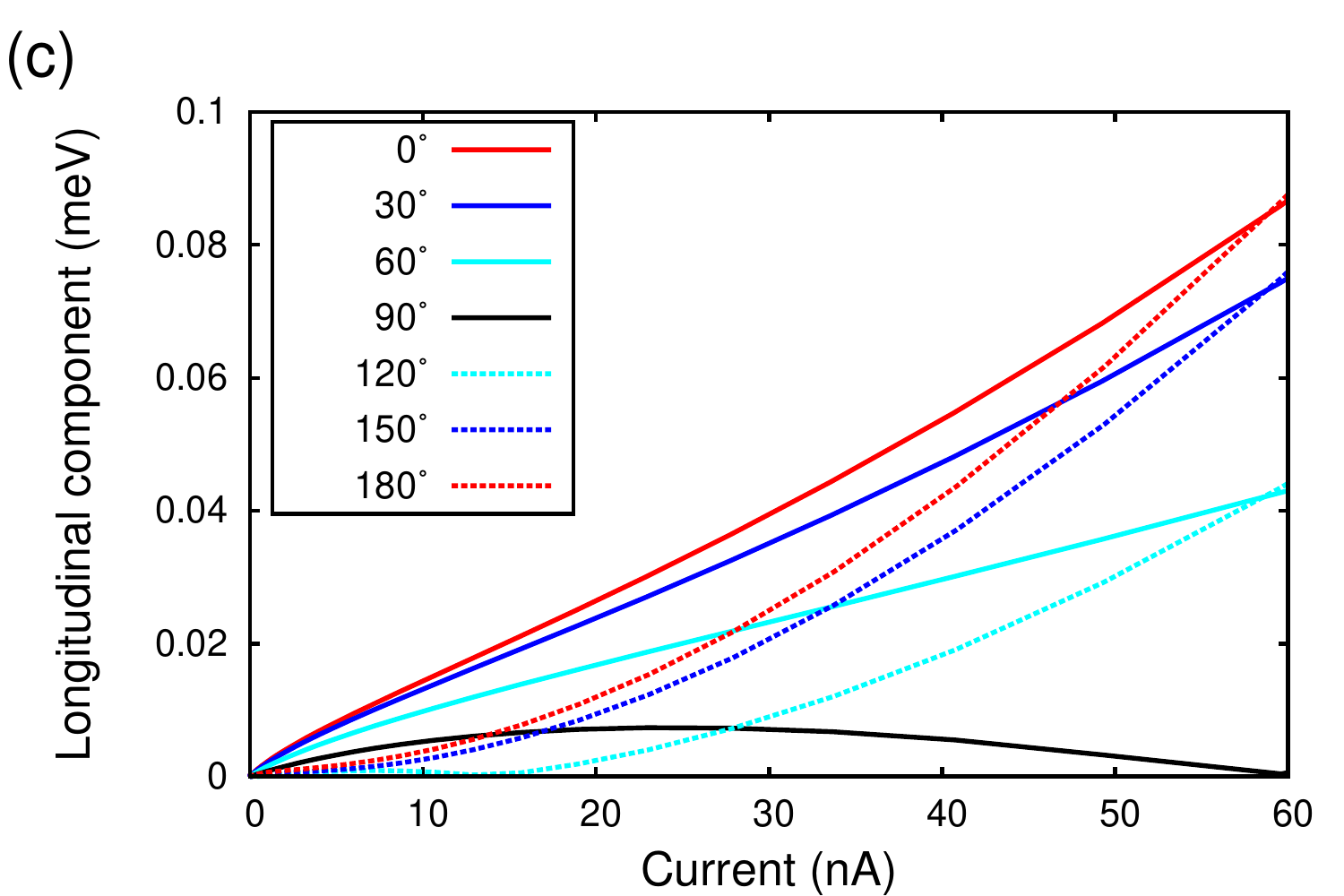}&
\includegraphics[width=0.5\textwidth,angle=0]{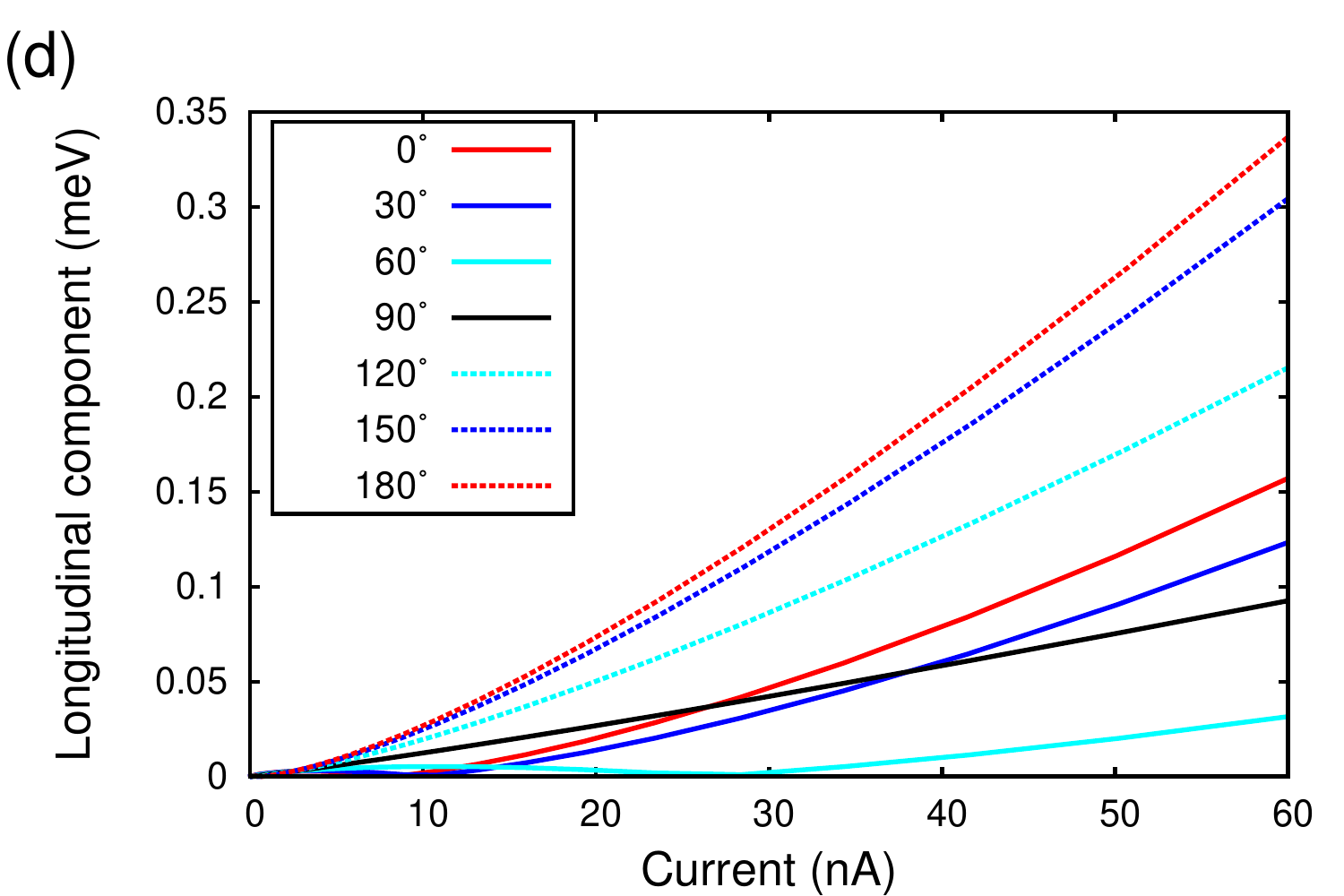}
\end{tabular}
\caption{\label{Fig3} (Color online) Magnitude of the spin transfer torque (a,b) and the longitudinal spin
current (c,d) as a function of the charge current above the top (a,c) and hollow (b,d) sites
of the Fe/W(110) surface in seven relative tip-sample magnetization orientations [$\phi$ between 0 and 90 degrees
in $15^{\circ}$ step in (a,b) because STT($\pi/2-\phi$)$\approx$STT($\pi/2+\phi$), and $\phi$ between 0 and 180 degrees in
$30^{\circ}$ step in (c,d)] at -0.2 V bias voltage.
}
\end{figure*}

Figure \ref{Fig3} shows the magnitude of the STT and the longitudinal spin current as a function of the charge
current above two different surface sites (top and hollow) of the Fe/W(110) surface. The bias voltage is fixed at -0.2 V and seven
different tip magnetization orientations are considered for the two spin transport quantities. The choice of this
bias voltage was motivated by the comparability of our calculation results with the experimental work of Krause {\it et al.}
\cite{krause11}.
Our first finding is that the torque-current curves deviate from the linear relationship reported for similar curves
of the experimentally determined modified activation energy barrier directly related to the STT in Ref.\ \cite{krause11}.
Our finding can be understood as follows. Let us assume that both the charge current and the STT are exponentially decaying with
increasing tip-sample distance $z$, i.e., $I(z)\propto\exp\{-\kappa_I z\}$ and $T(z)\propto\exp\{-\kappa_T z\}$. This assumption
is validated by fitting exponential functions to the calculated $I(z)$ and $T(z)$ curves (not shown), and we obtain $\kappa_I$
either 2.002 or 2.008 \AA$^{-1}$ and $\kappa_T$ either 1.748 or 1.768 \AA$^{-1}$ depending
on the lateral STM tip position (above the top and hollow surface site, respectively).
Due to the almost perfect $\sin(\phi)$-scaling of the torque-current curves, the decay constants do not depend
much on the relative tip-sample magnetization orientation ($\phi$). Note that an exponential $z$-decay of STT vector components
using a free-electron model was also found in Ref.\ \cite{wilczynski08}. The relationship $T(I)\propto I^{\kappa_T/\kappa_I}$ can
then directly be derived. This shows that the linear relationship between the STT and the charge current is ensured only if
$\kappa_T=\kappa_I$. To analyze our calculated $T(I)$ functions in more detail, Table \ref{Table1} reports fitted parameters
assuming power and linear function forms for $T(I)$. Based on the coefficients of determination ($R^2$) it is found that the power
function fit is always better than the linear fit. Thus, based on Table \ref{Table1}, we can conclude on the more
general power function behavior of the torque-current curves in Fig.\ \ref{Fig3}. The obtained $\kappa_T/\kappa_I$ is always found
to be smaller than one, which means that the decay constant of the STT ($\kappa_T$) is smaller than that of the charge current
($\kappa_I$), see also the range of explicitly fitted values above. This immediately implies that the STT efficiency
\cite{kalitsov13} (or normalized torque \cite{wilczynski08}) defined by the $T/I$ ratio can be enhanced at larger tip-sample
distances since $T/I\propto\exp\{(\kappa_I-\kappa_T)z\}$. We return to this later on. According to Table \ref{Table1} we found
$\kappa_T/\kappa_I$ either $\approx$0.873 or $\approx$0.881 above the top and hollow surface site, respectively.

\begin{table}[h]
\begin{centering}
\begin{tabular}{ccccc}
\hline
\hline
& \multicolumn{2}{c}{$T(I)\propto I^{\kappa_T/\kappa_I}$ fit} & \multicolumn{2}{c}{$T(I)=T_0+cI$ fit}\tabularnewline
\hline
& Fig.\ \ref{Fig3}(a) & Fig.\ \ref{Fig3}(b) & Fig.\ \ref{Fig3}(a) & Fig.\ \ref{Fig3}(b)\tabularnewline
$\phi$ [deg] & top & hollow & top & hollow\tabularnewline
\hline
& \multicolumn{2}{c}{$\kappa_T/\kappa_I$ ($R^2$)} & \multicolumn{2}{c}{$c$ [meV/$\mu$A] ($R^2$)}\tabularnewline
15 & 0.873 (0.999) & 0.881 (0.964) & 0.328 (0.979) & 0.934 (0.956)\tabularnewline
30 & 0.873 (0.999) & 0.881 (0.964) & 0.634 (0.979) & 1.804 (0.956)\tabularnewline
45 & 0.873 (0.999) & 0.881 (0.964) & 0.896 (0.979) & 2.549 (0.957)\tabularnewline
60 & 0.873 (0.999) & 0.880 (0.964) & 1.097 (0.979) & 3.118 (0.957)\tabularnewline
75 & 0.873 (0.999) & 0.880 (0.964) & 1.223 (0.979) & 3.473 (0.957)\tabularnewline
90 & 0.872 (0.999) & 0.880 (0.964) & 1.266 (0.979) & 3.590 (0.957)\tabularnewline
\hline
\hline
\end{tabular}
\par\end{centering}
\caption{\label{Table1} Fitted parameters on the torque-current curves in Fig.\ \ref{Fig3} (a,b) assuming power
and linear function forms for $T(I)$. The coefficients of determination ($R^2$) are also reported, which indicate the quality of
the fits.}
\end{table}

Although the fitted torque-current curves show a preference of the power function for $T(I)$, it is interesting to compare the
obtained slopes of the linear fits with the one reported in Ref.\ \cite{krause11}. Krause {\it et al.} calculated a slope of
$c=$ 1.5 meV/$\mu$A for the charge current dependence of the modified activation energy barrier, $\Delta E(I)=cI$. Our fitted $c$
slopes for $T(I)=T_0+cI$ vary between 0.328 and 3.590 meV/$\mu$A, thus these values are found well in the same
order of magnitude as in the experiment. The quantitative difference can be attributed to the different tip material used, Fe(001)
tip in our case and Cr-coated W tip in the experiment, or to the variation of $\phi$:
$c(\phi)\approx\sin(\phi)c(90^{\circ})$ in Table \ref{Table1}. Although the relative tip-sample magnetization orientation was
not explicitly reported in Ref.\ \cite{krause11}, based on our linear fit results we suggest that the
experimental STM setup corresponds to $\phi\approx 25^{\circ}$ [=$\arcsin(1.5/3.59)$] assuming the tip apex above
the surface hollow site to obtain the experimentally determined $c=$ 1.5 meV/$\mu$A. Note that the experimental value cannot be
reached with a tip apex above the surface top site, where the maximal slope of $c=$ 1.266 meV/$\mu$A has been calculated for
$\phi=90^{\circ}$.

In Fig.\ \ref{Fig3} (a,b) we find that the STT is generally larger above the hollow rather than above the top
surface position. This finding is important in the close-contact regime only since the visible region of Fig.\ \ref{Fig3}
corresponds to the $I\approx$ 1-60 nA current range and a tip-sample distance range of $z\approx$ 5-3 \AA. Note that in this $z$
range the charge current at a fixed $z$ position (and at -0.2 V bias) is larger above the hollow than above the top position,
i.e., $\Delta I(z,$-0.2 V$)=I^{top}(z,$-0.2 V$)-I^{hollow}(z,$-0.2 V$)<0$, see Fig.\ \ref{Fig2}. Since the STT decays slower than
the charge current ($\kappa_T/\kappa_I<1$), the above explain the generally larger STT above the hollow surface site. This is,
in effect, due to the largest orbital-dependent contributions from nearest surface atoms at the STM tip position in the 3D-WKB
picture.

Fig.\ \ref{Fig3} (c,d) show the longitudinal spin current ($T^{long}$) as a function of the charge current above
top and hollow surface sites of the Fe/W(110) surface at different $\phi$ angles. It is clear from Eq.\ (\ref{Eq_current_decomp})
that $T^{long}$ does not obey the angular symmetry relation of the STT, and qualitatively different $T^{long}-I$ relationships are
obtained below and above $\phi=90^{\circ}$. The power law behavior of $T^{long}(I)$ seems to be preserved. Similarly as found for
the STT, $T^{long}$ is generally larger above the hollow rather than above the top position. Moreover, we find that $T^{long}$ is
comparable in size with the STT and even larger close to the collinear alignment of $\mathbf{e}_S$ and $\mathbf{e}_T$. The
longitudinal spin current, however, does not contribute to the current induced magnetization rotation caused by the STT vector.

\begin{figure*}
\includegraphics[width=0.5\textwidth,angle=0]{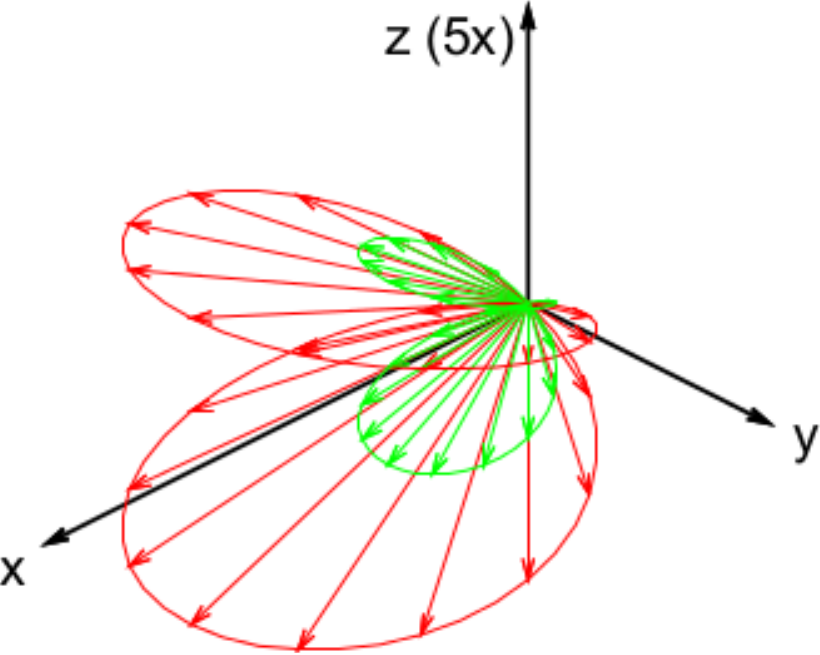}
\caption{\label{Fig4} (Color online) Spin transfer torque vectors 5 \AA\;above the top (red) and hollow
(green) sites of the Fe/W(110) surface in 24 relative tip-sample magnetization orientations
($\phi$ between 0 and 360 degrees in $15^{\circ}$ step) at -0.2 V bias voltage.
}
\end{figure*}

To get more insight to the STT, we investigate its vector character. Figure \ref{Fig4} shows calculated STT vectors at -0.2 V bias
voltage and 5 \AA\;above the top and hollow sites of the Fe/W(110) surface considering 24 relative tip-sample
magnetization orientations. It is clearly seen that the end-points of the STT vectors follow a continuous curve in
space, and the out-of-plane (z) components obey a $\sin(\phi)$ dependence. The in-plane components rotate in the $x-y$ plane and
their size follows a $|\sin(\phi)|$ function. Increasing the tip-sample distance, we find that the curves
formed by the $\phi$-dependent STT vectors above the top and hollow sites converge to the same curve, reaching
equivalence above $z=8$ \AA\;(not shown). These findings prove that the STT vectors depend on the lateral
(and vertical) STM tip position even above a clean (defect-free) planar ferromagnetic surface in a certain
tip-sample distance regime (3 \AA\;$<z<$ 8 \AA\;in our case).

For designing real devices using the STT for current induced magnetization switching, it is ultimately important to maximize the
STT efficiency. We adopt the definition of the STT efficiency as reported in Ref.\ \cite{kalitsov13}: in-plane torque ratio =
$|\mathbf{T}^{\parallel}|/I$, out-of-plane torque ratio = $|\mathbf{T}^{\perp}|/I$. In the following we analyze the STT efficiency
and the longitudinal spin current ratio ($T^{long}/I$) depending on the tip-sample distance, the relative
tip-sample magnetization orientation and the bias voltage.

\begin{figure*}
\begin{tabular}{cc}
\includegraphics[width=0.5\textwidth,angle=0]{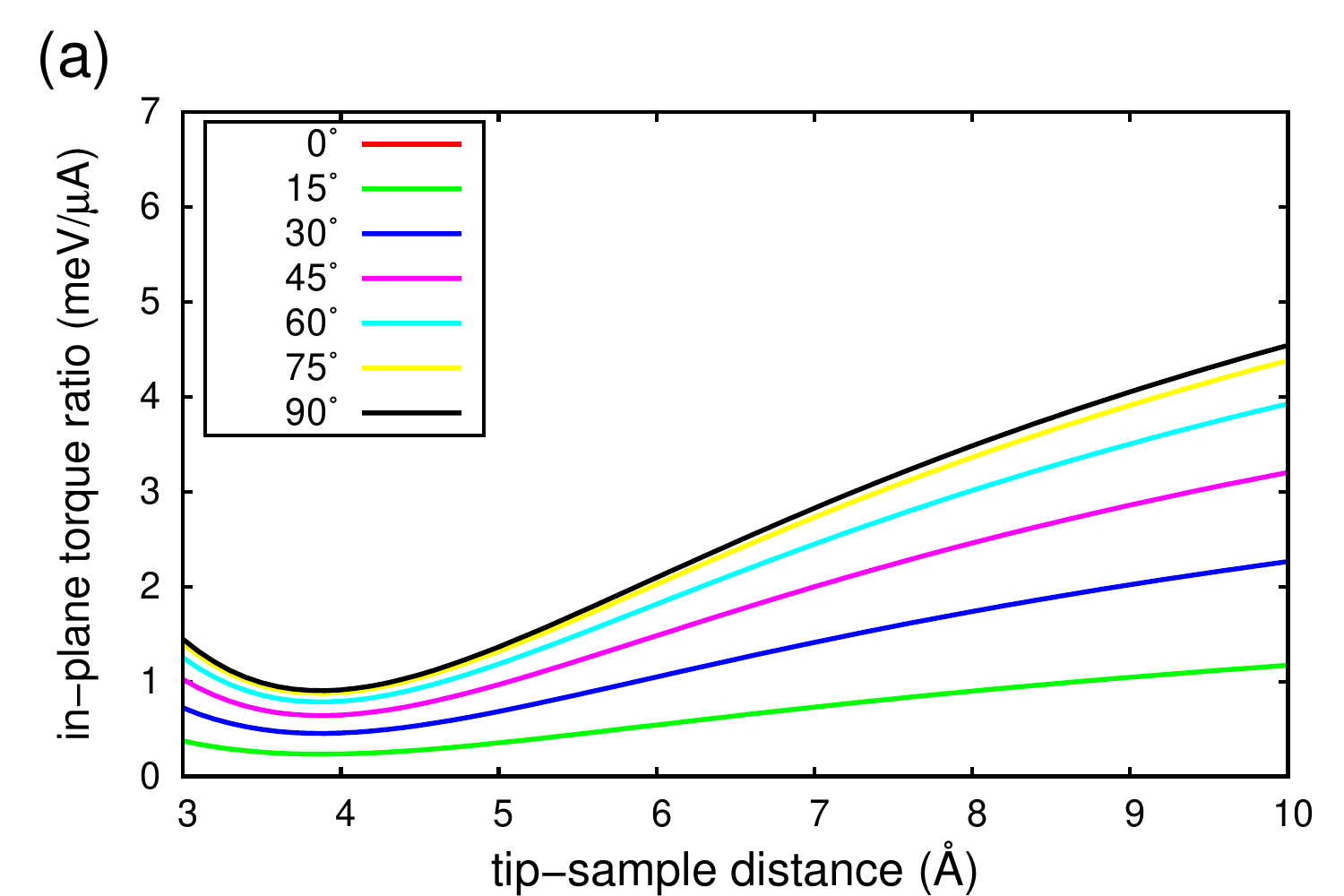}&
\includegraphics[width=0.5\textwidth,angle=0]{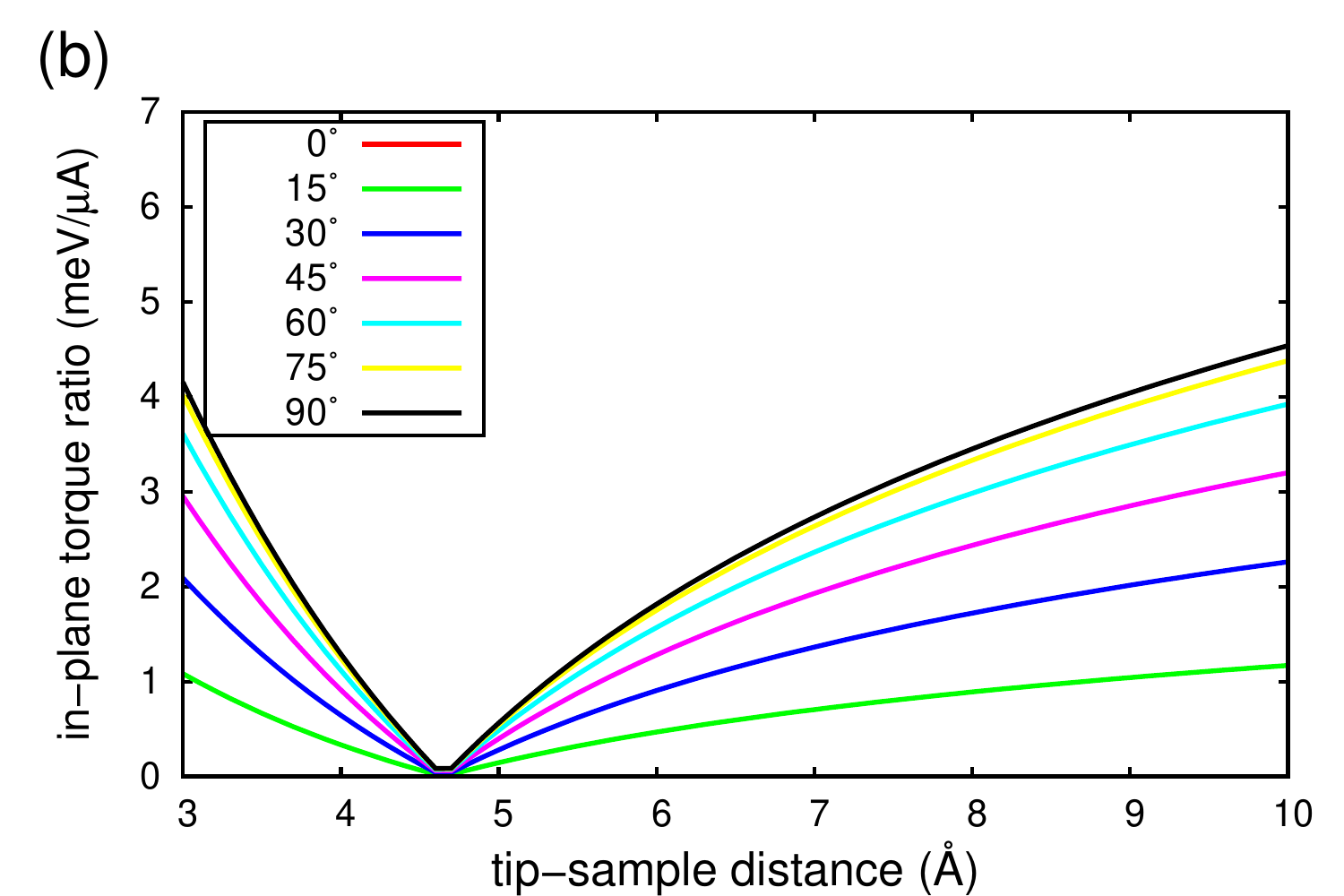}\tabularnewline
\includegraphics[width=0.5\textwidth,angle=0]{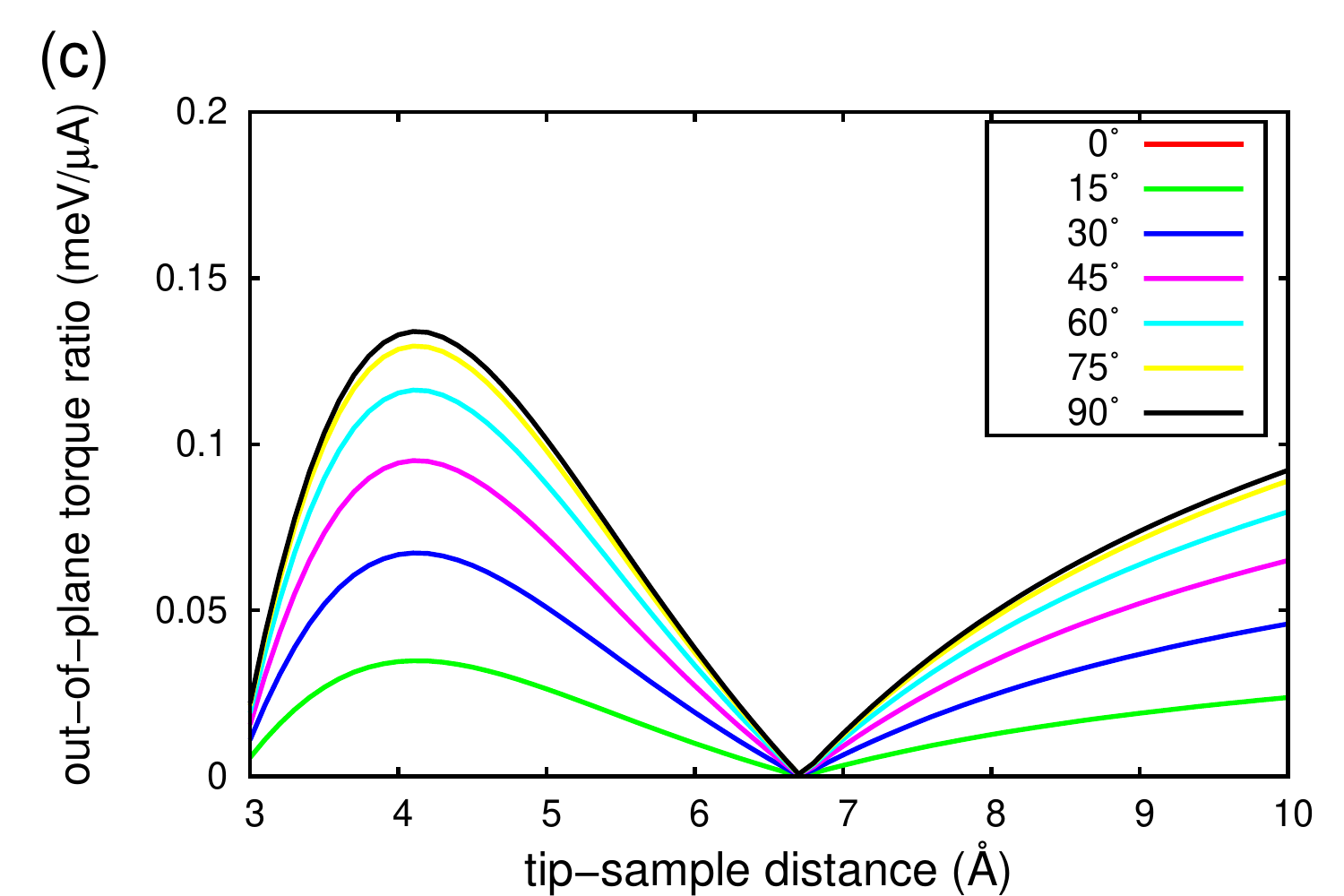}&
\includegraphics[width=0.5\textwidth,angle=0]{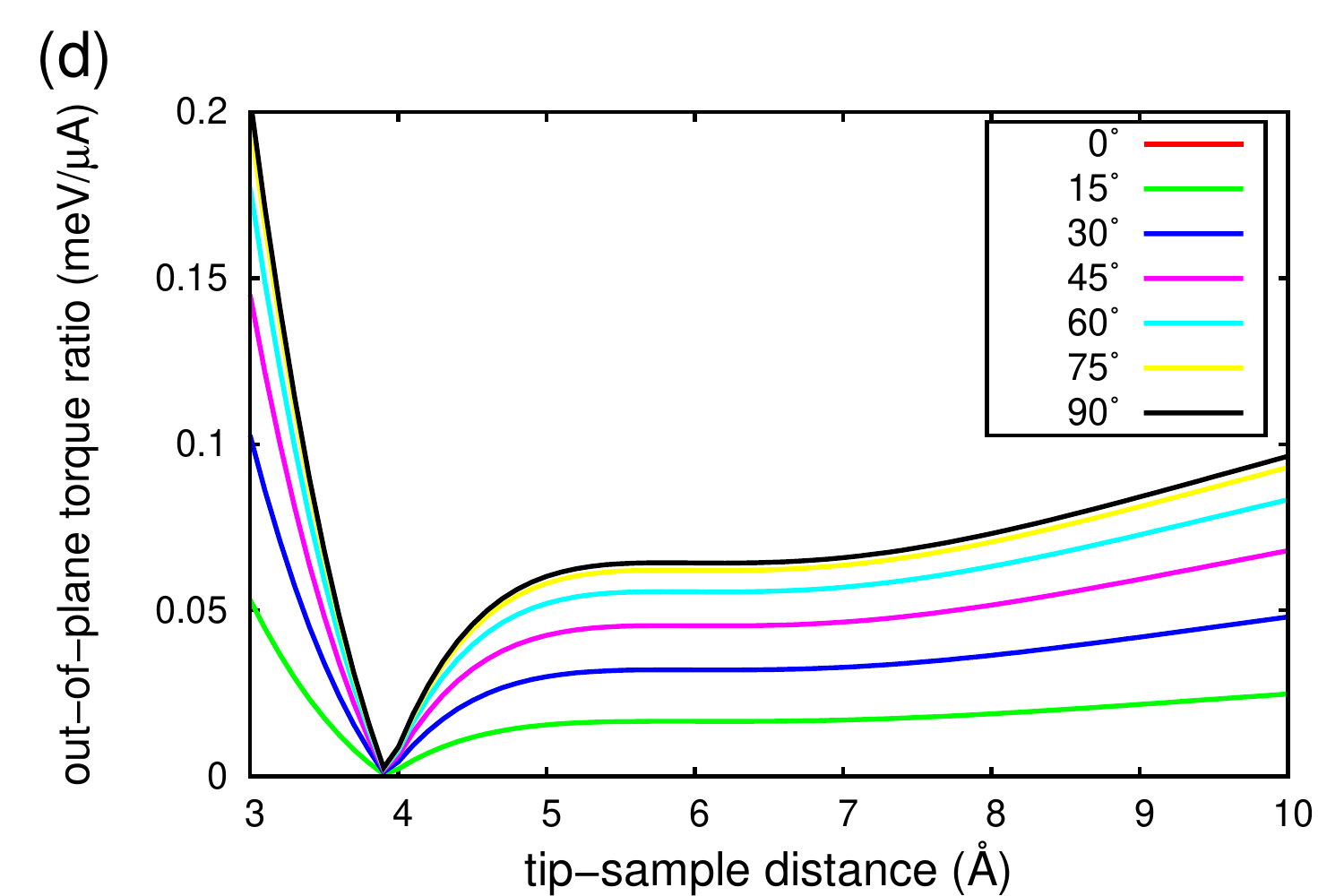}\tabularnewline
\includegraphics[width=0.5\textwidth,angle=0]{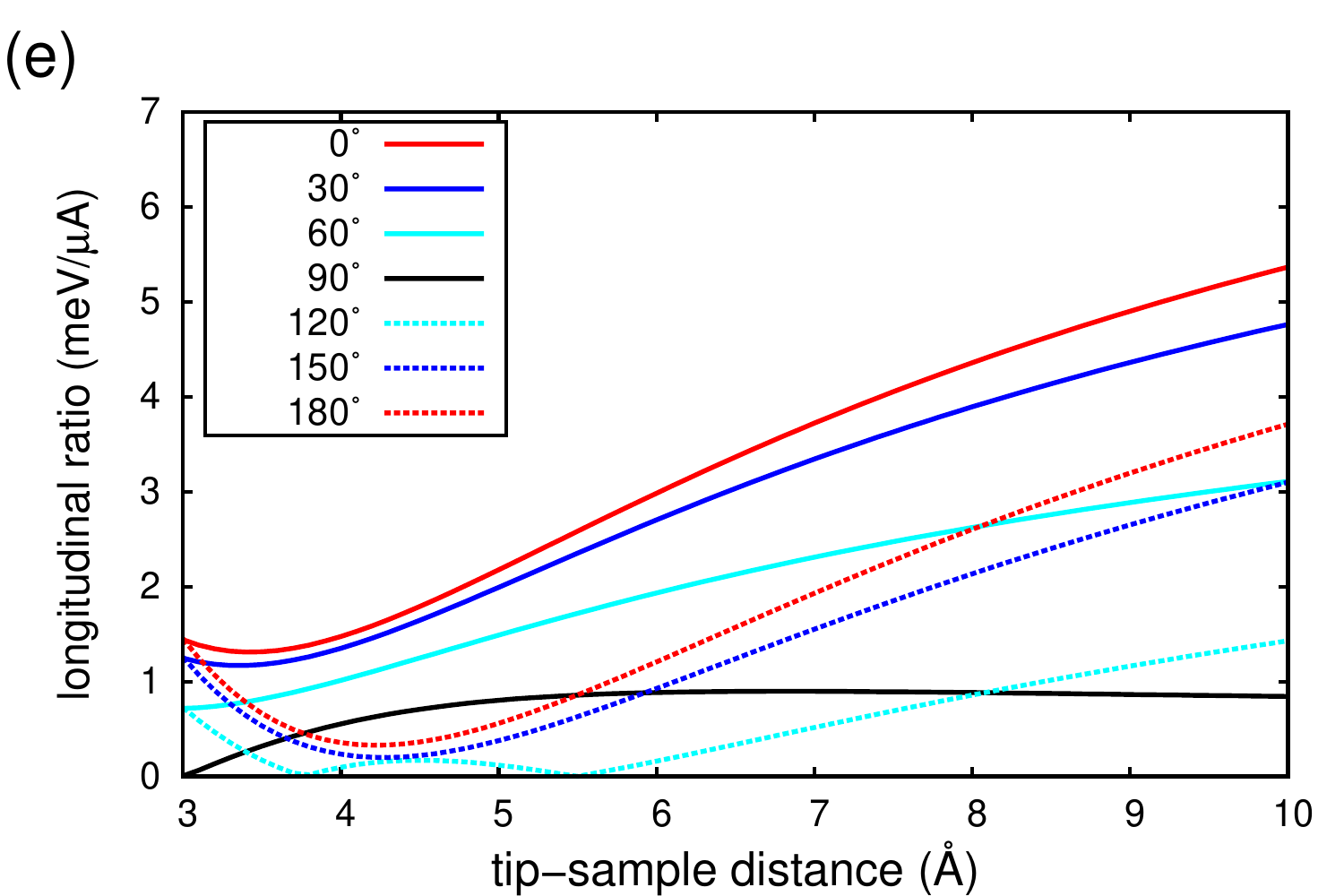}&
\includegraphics[width=0.5\textwidth,angle=0]{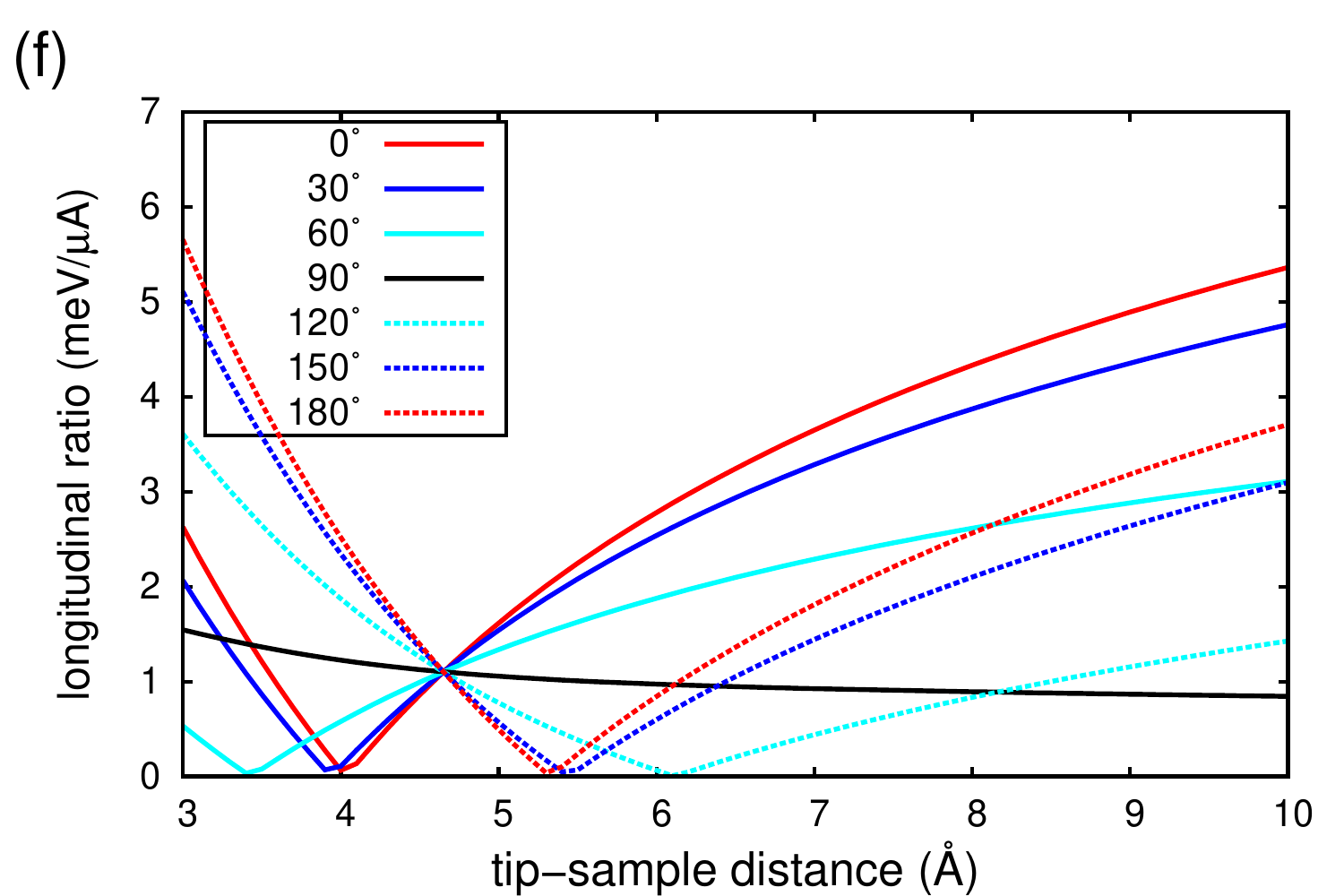}
\end{tabular}
\caption{\label{Fig5} (Color online) STT efficiency: Ratios between spin transfer torque components ($T^{\parallel,\perp}$) and
longitudinal spin current ($T^{long}$) and the charge current ($I$) above two surface positions of the Fe/W(110)
surface in seven relative tip-sample magnetization orientations [$\phi$ between 0 and 90 degrees in $15^{\circ}$
step in (a-d), and $\phi$ between 0 and 180 degrees in $30^{\circ}$ step in (e,f)] at -0.2 V bias voltage as the function of the
tip-sample distance:
(a) in-plane torque ratio ($T^{\parallel}/I$) above the top position;
(b) in-plane torque ratio ($T^{\parallel}/I$) above the hollow position;
(c) out-of-plane torque ratio ($T^{\perp}/I$) above the top position;
(d) out-of-plane torque ratio ($T^{\perp}/I$) above the hollow position;
(e) longitudinal ratio ($T^{long}/I$) above the top position;
(f) longitudinal ratio ($T^{long}/I$) above the hollow position.
}
\end{figure*}

Figure \ref{Fig5} shows the tip-sample-distance-dependent in-plane and out-of-plane STT efficiencies and the
longitudinal spin current ratio calculated at -0.2 V bias voltage above two surface positions of the Fe/W(110) surface in seven
relative tip-sample magnetization orientations each. We find that the $z$-dependence of the in-plane and
out-of-plane STT efficiencies differs considerably, and the in-plane component is always larger by at least a
factor of four than the out-of-plane one, in agreement with Refs.\ \cite{theodonis06,heiliger08}. This means that the magnitude
of the STT is, in effect, determined by the in-plane component at this bias voltage, similarly found in metallic spin valves
\cite{ralph08}. We also find considerable differences between the $z$-dependent STT efficiencies above the top and hollow surface
sites of Fe/W(110). The in-plane STT efficiencies (and thus the magnitudes of the STT) converge to the same $\phi$-dependent
curves above $z=$ 8 \AA\;for both surface sites. We recover the already mentioned exponentially increasing STT efficiency
[$T/I\propto\exp\{(\kappa_I-\kappa_T)z\}$] at large tip-sample distances (above $z=$ 8 \AA).
Note that in symmetric tunnel junctions an opposite type of $z$-decay (faster STT than charge current decay) can occur,
which results in an exponentially vanishing STT efficiency as the tip-sample distance (barrier thickness) increases
\cite{wilczynski08}.
In our asymmetric STM junction, we find the maximal STT efficiency values at the investigated upper boundary of $z=$ 10 \AA\;for
all considered $\phi$ values above both surface sites [Fig.\ \ref{Fig5}(a) and (b)].
Interestingly, we also observe large STT efficiency at small tip-sample distances ($z<4$ \AA)
above the surface hollow site [Fig.\ \ref{Fig5}(b)].

For the out-of-plane STT efficiency we find that maximal values can be obtained at small tip-sample distances: The local maximum
is found at $z=$ 4.1 \AA\;in Fig.\ \ref{Fig5}(c) and at $z=$ 3 \AA\;in Fig.\ \ref{Fig5}(d). However, there are
local minima at $z=$ 6.7 \AA\;and $z=$ 3.9 \AA, respectively, in the latter case quite close to the local maximum.
Note that the almost perfect $\sin(\phi)$-scaling of the STT efficiency curves in Fig.\ \ref{Fig5}(a)-(d) is
recovered. The longitudinal spin current ratios ($T^{long}/I$) shown in Fig.\ \ref{Fig5}(e,f) are comparable in size with the
in-plane STT efficiencies and they show complex $z$- and $\phi$-dependences resulting from a complex interplay of
orbital-dependent transmission and electronic structure effects, see Eq.\ (\ref{Eq_current_decomp}).

\begin{figure*}
\includegraphics[width=1.0\textwidth,angle=0]{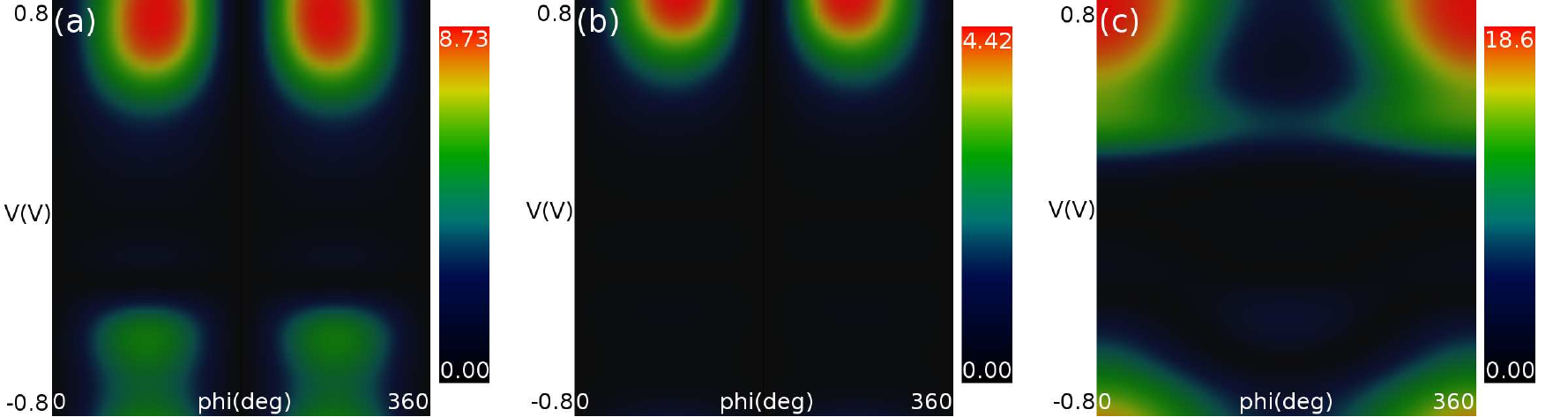}
\caption{\label{Fig6} (Color online) STT efficiency: Ratios between spin transfer torque components ($T^{\parallel,\perp}$) and
longitudinal spin current ($T^{long}$) and the charge current ($I$) 5 \AA\;above the top surface
position of the Fe/W(110) surface as the function of the relative tip-sample magnetization orientation ($\phi$) and the bias
voltage ($V$):
(a) in-plane torque ratio ($T^{\parallel}/I$);
(b) out-of-plane torque ratio ($T^{\perp}/I$);
(c) longitudinal ratio ($T^{long}/I$).
The values in the scale bars are given in units of meV/$\mu$A.
}
\end{figure*}

Finally, we consider the possibility of tuning the STT efficiency by the applied bias voltage in the tunnel junction.
Figure \ref{Fig6} shows in-plane and out-of-plane STT efficiency maps and a longitudinal spin current ratio map
5 \AA\;above the top surface position of the Fe/W(110) surface as the function of the relative tip-sample
magnetization orientation and the bias voltage. We find that the obtained STT efficiency maps are qualitatively similar above
the top and hollow surface sites with maximal values found close to $\phi=90^{\circ}$ and $270^{\circ}$, which
is also true for the studied $z$ range of [3 \AA, 10 \AA] (not shown).
However, the longitudinal spin current ratio map shows a totally different angular dependence with maximal values
close to $\phi=0^{\circ}$.

We identify $|V|>0.4$ V bias voltage ranges for highly enhanced STT efficiencies.
A bias-asymmetry in the favorable torque efficiencies is evident in Fig.\ \ref{Fig6}(a,b) as reported in other
asymmetric tunnel junctions \cite{kalitsov13}. Asymmetric bias dependence of the STT has been reported earlier
\cite{slonczewski05,theodonis06}. Moreover, we find that the angular dependence of the STT efficiencies is
symmetric to $\phi=90^{\circ}$ at low bias but asymmetric at high bias. A similar behavior has been explained by a resonant
transmission channel in Ref.\ \cite{jia11}. In our 3D-WKB model, the bias- and angle-asymmetry are related to the integrated
electronic structures in the bias window, where the transport quantities involving the magnetic PDOS ($m_S$ or $m_T$) can have a
nonmonotonic behavior and can even change sign depending on the bias voltage, see Eq.\ (\ref{Eq_current_decomp}). Such a behavior
is reported for the spin-polarized part of the charge current in Ref.\ \cite{palotas13contrast}. Although the
torque-current relationship keeps its general power law dependence, the $\kappa_T(V)/\kappa_I(V)$ vacuum-decay ratio
will surely be bias dependent and can be different than reported for -0.2 V bias.

We obtain a maximal STT efficiency of 1.369 meV/$\mu$A at the experimentally used bias voltage of -0.2 V \cite{krause11} at
5 \AA\;tip-sample distance, whereas the linear fit in Table \ref{Table1} gave a maximum value of 1.266 meV/$\mu$A, both at
$\phi=90^{\circ}$. The direct calculation of the STT efficiency thus provides a better agreement with the experimentally
determined 1.5 meV/$\mu$A than the linear fit, and tuning the tip-sample distance would result in a perfect agreement.
This finding already highlights the problem of the linear fit of the torque-current curves. This problem is even more striking
when placing the STM tip above the hollow surface position, where we obtained 1.5 meV/$\mu$A at $\phi\approx 25^{\circ}$ based
on the linear fit but a maximal value of 0.566 meV/$\mu$A is found at $\phi=90^{\circ}$ and 5 \AA\;in the direct STT efficiency
calculation. Again, by tuning the tip-sample distance a perfect agreement with the experimental value could be achieved.
Nevertheless, our most important finding is that the STT efficiency can be enhanced by about a factor of seven
from 1.369 (at -0.2 V and $\phi=90^{\circ}$) to 9.706 meV/$\mu$A (at 0.8 V and $\phi=100^{\circ}$) by changing the bias voltage
and $\phi$ compared to the experimentally used voltage of -0.2 V \cite{krause11}.

All reported characteristics of the tip-sample-distance-dependent (determined by mostly geometric effect) and the
bias-voltage-dependent (determined by mostly electronic structure effect) STT efficiencies are due to the complex interplay of the
orbital-dependent tunneling (determined by the geometry of the STM junction) and the involved spin-polarized electronic
structures of the sample surface and the STM tip. We anticipate that most of the reported effects can generally be found and
utilized in other magnetic STM junctions as well. Moreover, we suggest that by changing the tip material the bias-dependent
STT efficiencies could be further enhanced. This can be a topic of future research.

\section{Conclusions}
\label{sec_conc}

We have introduced a method for a combined calculation of charge and vector spin transport of elastically tunneling electrons in
magnetic scanning tunneling microscopy (STM). The method is based on the three-dimensional Wentzel-Kramers-Brillouin (3D-WKB)
approach combined with electronic structure calculations using first principles density functional theory.
Employing the model, we have analyzed the STM contrast inversion of the charge current above the Fe/W(110) surface depending on
the bias voltage, tip-sample distance and relative magnetization orientation between the sample and an iron tip.
We have demonstrated the tunability of the atomic contrast inversion in STM images depending on the contributing spin channels.
For the spin transfer torque (STT) vector we have found that its in-plane component is generally larger than the out-of-plane
component, and we identified a longitudinal spin current component, which, however, does not contribute to the
torque. We have shown that the STT vectors depend on the lateral and vertical STM tip positions even above a clean (defect-free)
planar ferromagnetic surface. Our results suggest that the torque-current relationship in magnetic STM junctions follows the
power law rather than a linear function. Consequently, we have shown that the ratio between the STT and the spin-polarized charge
current is not constant, and more importantly, it can be tuned by the bias voltage, tip-sample distance and magnetization rotation.
We found that the STT efficiency can be enhanced by about a factor of seven by selecting a proper bias voltage.
Thus, we demonstrated the possible enhancement of the STT efficiency in magnetic STM junctions, which can be exploited in
technological applications. We discussed our results in view of the indirect measurement of the STT above the Fe/W(110) surface
reported by Krause {\it et al.} \cite{krause11}. Our presented work is expected to inspire future research on high resolution
vector spin transport characteristics in magnetic STM junctions.

\section{Acknowledgments}

Financial support of the SASPRO Fellowship of the Slovak Academy of Sciences (project no.\ 1239/02/01), the Hungarian State
E\"otv\"os Fellowship and the National Research, Development and Innovation Office of Hungary under project no.\ K115575
is gratefully acknowledged.

\appendix
\section{In-plane torque vectors}

Here, we prove the formal equivalence of the in-plane torque vectors in Eq.\ (\ref{Eq_current_decomp}) with Eq.\ (6) of
Ref.\ \cite{theodonis06}. According to Theodonis {\it et al.}, the in-plane torque vector acting on $\mathbf{M}_2$ is directly
obtained from the longitudinal spin currents, and can be written using our notation as
\begin{equation}
\mathbf{T}^{\parallel,1\rightarrow 2}=\mathbf{T}^{2\parallel}=\frac{T^{long-2}(\phi=180^{\circ})-T^{long-2}(\phi=0^{\circ})}{2}\mathbf{M}_2\times(\mathbf{M}_1\times\mathbf{M}_2).
\label{Eq_theodonis}
\end{equation}
Using the longitudinal spin currents from Eq.\ (\ref{Eq_current_decomp}),
\begin{eqnarray}
T^{long-S}(\phi=0^{\circ})=T^{long-T}(\phi=0^{\circ})&\propto&n_S m_T + m_S n_T,\nonumber\\
T^{long-S}(\phi=180^{\circ})&\propto&m_S n_T - n_S m_T,\nonumber\\
T^{long-T}(\phi=180^{\circ})&\propto&n_S m_T - m_S n_T
\end{eqnarray}
in the two possible transport directions (tip $\rightarrow$ sample at $V>0$ and sample $\rightarrow$ tip at $V<0$) and the
identity,
$\mathbf{a}\times(\mathbf{b}\times\mathbf{c})=\mathbf{b}(\mathbf{a}\cdot\mathbf{c})-\mathbf{c}(\mathbf{a}\cdot\mathbf{b})$,
the in-plane torque vectors following Eq.\ (\ref{Eq_theodonis}) can be obtained as
\begin{eqnarray}
\mathbf{T}^{\parallel,T\rightarrow S}=\frac{T^{long-S}(180^{\circ})-T^{long-S}(0^{\circ})}{2}\mathbf{e}_S\times(\mathbf{e}_T\times\mathbf{e}_S)&\propto&-n_S m_T \left[\mathbf{e}_T-\mathbf{e}_S\cos(\phi)\right],\nonumber\\
\mathbf{T}^{\parallel,S\rightarrow T}=\frac{T^{long-T}(180^{\circ})-T^{long-T}(0^{\circ})}{2}\mathbf{e}_T\times(\mathbf{e}_S\times\mathbf{e}_T)&\propto&-m_S n_T \left[\mathbf{e}_S-\mathbf{e}_T\cos(\phi)\right].
\end{eqnarray}
These correspond to the in-plane torque formulas in Eq.\ (\ref{Eq_current_decomp}):
\begin{eqnarray}
\mathbf{T}^{S\parallel}&\propto&n_S m_T \left[\mathbf{e}_T-\mathbf{e}_S\cos(\phi)\right],\nonumber\\
\mathbf{T}^{T\parallel}&\propto&m_S n_T \left[\mathbf{e}_S-\mathbf{e}_T\cos(\phi)\right].
\end{eqnarray}

\end{document}